\documentclass[preprint]{ptephy_v1}

\preprintnumber{XXXX-XXXX} 
\usepackage{hyperref}

\usepackage{graphicx,color}

\usepackage{amsmath,amssymb}
\usepackage{url}
\usepackage{epstopdf}
\newcommand{\hs}{\hspace*{0.5cm}}

\newcommand{\be}{\begin{equation}}
	\newcommand{\ee}{\end{equation}}
\newcommand{\bea}{\begin{eqnarray}}
	\newcommand{\eea}{\end{eqnarray}}
\newcommand{\nn}{\nonumber}
\newcommand{\crn}{\nonumber \\}

\newcommand{\al}{\alpha}
\newcommand{\la}{\lambda}
\newcommand{\bet}{\beta}

\newcommand{\om}{\omega}
\newcommand{\pa}{\partial}
\newcommand{\fr}{\frac}

\newcommand{\bc}{\begin{center}}
	\newcommand{\ec}{\end{center}}

\newcommand{\ep}{\epsilon}

\newcommand{\ph}{\phi}

\newcommand {\ba}{\begin{array}}
	\newcommand {\ea}{\end{array}}
\newcommand{\ben}{\begin{enumerate}}
	\newcommand{\een}{\end{enumerate}}

\usepackage{bm}
\usepackage{dcolumn}

\begin{document}

\title{$(g-2)_{e,\mu}$ and decays $e_b\to e_a\gamma$  in a  $ \mbox{SU}(4)_L \otimes \mbox{U}(1)_X$ model with inverse seesaw neutrinos}


\author{N. H. Thao}
\affil{Department of Physics, Hanoi Pedagogical University 2, no 32 Nguyen Van Linh, Phuc Yen, Vinh Phuc, Vietnam 
\email{nguyenhuythao@hpu2.edu.vn}
}

\author[2,3]{D. T. Binh}
\affil[2]{Institute of Theoretical and Applied Research, Duy Tan University, Hanoi, Vietnam }
\affil[3]{Faculty of Natural Science, Duy Tan University, Da Nang, Vietnam \email{dinhthanhbinh3@duytan.edu.vn} }

\author[4]{T.~T.~Hong}
\affil[4]{An Giang University, VNU - HCM, Ung Van Khiem Street,
	Long Xuyen, An Giang, Vietnam
	  \email{tthong@agu.edu.vn}
} 

\author[5,6]{L. T. Hue }
\affil[5]{Subatomic Physics Research Group, Science and Technology Advanced Institute, Van Lang University, Ho Chi Minh City, Vietnam }
\affil[6]{Faculty of Applied Technology, School of Technology, Van Lang University, Ho Chi Minh City, Vietnam  \email{lethohue@vlu.edu.vn}}

\author[7]{D. P. Khoi  
}
\affil[7]{Department of Physics, Vinh University, 182 Le Duan, Vinh City, Nghe An, Vietnam \email{khoidp@vinhuni.edu.vn} }



\begin{abstract}%
We will show that  the 3-4-1 model with heavy right-handed neutrinos can explain the recent experimental data of $(g-2)_{e, \mu}$ anomalies of charged leptons  and neutrino oscillations through the inverse seesaw mechanism. In addition, the model can predict large lepton flavor violating decay rates  $\mu \to e\gamma$ and $\tau \to \mu\gamma, e\gamma$  up to the recent experimental sensitivities.
\end{abstract}


\maketitle
\allowdisplaybreaks
\section{Introduction}
\label{sec:intro}
The 3-4-1 model with right-handed neutrinos (341RHN) was discussed in Ref.~\cite{Foot:1994ym,Pisano:1994tf} as a natural extension that  new right-handed neutrinos are assigned into the same left-handed lepton quadruplets.  For a complete  study of the highest possible gauge   symmetry in the electroweak sector \cite{Voloshin:1987qy}, various 3-4-1 extensions were introduced  with different electric charges of new exotic leptons \cite{Ponce:2003uu, Sanchez:2004uf, Ponce:2006vw, Sanchez:2008qv, Riazuddin:2008yx, Nam:2009tr,  Long:2016lmj, Palacio:2016mam, Djouala:2019hhn}. It was proved that   these  original  3-4-1 models  can not explain the recent data of the $(g-2)$ anomaly of muon  unless they must be extended  such as adding new inert scalars \cite{Cogollo:2020nrc}.    A solution applied to the  3-3-1 models \cite{Hue:2021zyw}, that adding  new singly charged Higgs bosons and inverse seesaw (ISS) neutrinos, is  another viable  approach.  In this work, we will investigate the possibility of whether this approach can work in the 3-4-1 model framework, which can  accommodate  data of  charged lepton anomalies $a_{e_a}\equiv (g-2)_{e_a}/2$, neutrino oscillation data,  and  the recent bounds of the lepton flavor violating decays of charged leptons (cLFV) $e_b\to e_a \gamma$.   The  explanations of neutrino oscillation data were mentioned previously in various 3-4-1 models, including the ISS mechanism \cite{Palcu:2009uk, Palcu:2015ica, Palacio:2016mam}, but they did not relate to the $(g-2)$ data and cLFV decays.  In the ISS models, new gauge contributions to $(g-2)$ are suppressed \cite{Pinheiro:2021mps, Palcu:2015ica}, especially the 3-3-1 and 3-4-1 models, because the new gauge bosons must be heavy to guarantee the recent lower bounds from experimental searches \cite{Lee:2014kna}.    Therefore,  we will study  the appearance of  new singly charged Higgs bosons and their mixing with the $SU(4)_L$ ones  that results in   large chirally-enhanced one-loop Higgs contributions to $a_{e_a}$ enough to be consistent with experiments \cite{Crivellin:2018qmi}. On the other hand, the ISS mechanism may lead to large one-loop contributions to cLFV rates. In this work, the numerical investigations  to determine the allowed regions of parameter space satisfying all experimental constraints of $(g-2)$ anomaly and cLFV decays $e_b\to e_a \gamma$ will be discussed precisely in the 3-4-1 framework.   Besides many available  models beyond the standard model (BSM) \cite{Han:2018znu, Endo:2019bcj, Abdullah:2019ofw,  Li:2020dbg, DelleRose:2020oaa, Bigaran:2020jil, Botella:2020xzf, Bharadwaj:2021tgp, Han:2021gfu, Arbelaez:2020rbq, Chun:2020uzw, Chen:2020tfr, Dutta:2020scq, Wang:2022yhm, Hernandez:2021tii, Hernandez:2021xet, Li:2022zap, Botella:2022rte, Kriewald:2022erk, Barman:2021xeq, Dermisek:2022hgh, Chowdhury:2022jde, Chen:2023eof} under various experimental data including the  $(g-2)_{e,\mu}$ anomalies,  our work will confirm the reality of the 3-4-1 models. 

The latest experimental measurement for muon anomaly  $a_{\mu}$ has been reported from    the combination of  the two experimental  results at 
Fermilab~\cite{Muong-2:2021ojo} and  Brookhaven National Laboratory (BNL) E82~\cite{Muong-2:2006rrc}:  $a^{\mathrm{exp}}_{\mu}=116592061(41)\times 10^{-11}$. It   leads  to a standard deviation of 4.2 $\sigma$ (standard deviation)  from the Standard Model  (SM) prediction, namely 
\begin{equation}\label{eq_damu}
	\Delta a^{\mathrm{NP}}_{\mu}\equiv  a^{\mathrm{exp}}_{\mu} -a^{\mathrm{SM}}_{\mu} =\left(2.51\pm 0.59 \right) \times 10^{-9},
\end{equation} 
where  $a^{\mathrm{SM}}_{\mu}= 116591810(43)\times 10^{-11}$ is the  SM prediction \cite{Aoyama:2020ynm} combined from various different contributions  based on the dispersion approach \cite{Davier:2010nc, Danilkin:2016hnh,  Davier:2017zfy, Keshavarzi:2018mgv, Colangelo:2018mtw, Hoferichter:2019mqg, Davier:2019can, Keshavarzi:2019abf, Kurz:2014wya, Melnikov:2003xd, Masjuan:2017tvw, Colangelo:2017fiz, Hoferichter:2018kwz, Gerardin:2019vio, Bijnens:2019ghy, Colangelo:2019uex, Colangelo:2014qya, Blum:2019ugy, Aoyama:2012wk, Aoyama:2019ryr, Czarnecki:2002nt, Gnendiger:2013pva, Pauk:2014rta, Jegerlehner:2017gek, Knecht:2018sci,Eichmann:2019bqf,Roig:2019reh}.   Another larger SM value  calculated using the lattice-QCD technique  implies a smaller value of $\Delta a^{\mathrm{NP}}_{\mu}$ than that given in Eq. \eqref{eq_damu} \cite{Borsanyi:2020mff}. In  this work we will accept the experimental constraint from Eq. \eqref{eq_damu} consisting of  both results. On the other hand, the recent experimental $a_e$ data  was  reported from different groups~\cite{Hanneke:2008tm, Parker:2018vye, Morel:2020dww, Fan:2022eto}, leading to the two inconsistent deviations  between experiments and the SM prediction \cite{Aoyama:2012wj,  Laporta:2017okg, Aoyama:2017uqe,  Terazawa:2018pdc, Volkov:2019phy, Gerardin:2020gpp}. In this work, we accept the following value\footnote{We thank the referee for providing us with the  experimental value of $a_e$ in Ref. \cite{Fan:2022eto}}:  
\begin{equation}\label{eq_dae}
	\Delta a^{\mathrm{NP}}_{e}\equiv  a^{\mathrm{exp}}_{e} -a^{\mathrm{SM}}_{e} = \left( 3.4\pm 1.6\right) \times 10^{-13},
\end{equation}  
where we have used the latest experimental result of $  a^{\mathrm{exp}}_{e}$ reported in Ref.  \cite{Fan:2022eto}, consistent with the 2008 result of the same group \cite{Hanneke:2008tm}, and the $a^{\mathrm{SM}}_{e}$ value reported in Ref. \cite{Morel:2020dww}, derived indirectly from the measurement of the fine-structure constant  $\alpha$ using the  Rb atoms, corresponding to the $2.1\sigma$ deviation. There is another $a^{\mathrm{SM}}_{e}$ value corresponding to the measurement of the fine-structure constant of Cs-133 atoms \cite{Parker:2018vye}, which is inconsistent with the earlier, namely $\Delta a^{\mathrm{NP}}_{e} = \left(-10.2\pm 2.6\right) \times 10^{-13}$, with the $3.9\sigma$ deviation. Both results can be explained in this work, see details later in the numerical investigation 

The ISS mechanism  may result in large values of not only $(g-2)_{e,\mu}$ but also Br$(e_b \to e_a \gamma)$, which  are constrained  by recent experiments \cite{BaBar:2009hkt,  MEG:2016leq, Belle:2021ysv}
\begin{align}
	\label{eq_eijkllimit}	
	&\mathrm{Br}(\tau\rightarrow \mu\gamma)<4.4 \times 10^{-8}, 
	\crn& \mathrm{Br}(\tau\rightarrow e\gamma) <3.3\times 10^{-8}, \;
	\crn &\mathrm{Br}(\mu\rightarrow e\gamma) < 4.2\times 10^{-13}. 
\end{align}
The future  sensitivities for  these decays are Br$(\mu \to e\gamma)<6\times 10^{-14}$ \cite{MEGII:2018kmf}, Br$(\tau\to e \gamma)< 9.0 \times 10^{-9}$, and Br$(\tau\to \mu \gamma)< 6.9 \times 10^{-9}$ \cite{Belle-II:2018jsg, Banerjee:2022xuw}.  Hence the correlations between $a_{e_a}$ and cLFV decays $e_b \to e_a \gamma$ may give new predictions on both of these kinds of processes, namely whether large $a_{e_a}$ will exclude the more strict experimental constraints of cLFV decays in the near future. 

Our paper is arranged as follows. Section \ref{sec_model} presents all ingredients of a 3-4-1 model to calculate the $(g-2)_{e_a}$ data and cLFV decays.   Section \ref{sec_lSSmodel} introduces the 341ISS model to accommodate the recent $(g-2)_{e_a}$ data. In this model,  the Yukawa Lagrangian and Higgs potential must respect a new global $U(1)_{\mathcal{L}}$ symmetry in order to guarantee the appearance of the ISS mechanism, the mixing between singly charged Higgs bosons,  and the Yukawa couplings resulting in large chirally-enhanced one-loop contributions to $(g-2)_{e_a}$ anomalies. Section \ref{sec_numerical} will present  detailed numerical results to determine the allowed regions of the parameter space that explain both experimental results of two $(g-2)_{e,\mu}$ anomalies  and  cLFV decays. Section \ref{sec_con} summarizes important results. 

\section{The model with Dirac active neutrinos}
\label{sec_model}
\subsection{\label{yukawa} Yukawa couplings and masses for fermions }
In this work, we will study the 3-4-1 model with heavy right-handed neutrinos and  new singly charged leptons assigned  in the three left-handed quadruplets  \cite{Sanchez:2004uf, Palacio:2016mam}.  This model is constructed based on the gauge symmetry $SU(3)_c\times SU(4)_L\times U(1)_X$, implying 16 electroweak (EW) gauge bosons. In addition, there are 4 neutral gauge bosons corresponding to the 4 diagonal generators of the EW group. Normally, the EW group is assumed to break to the final electric group through the following pattern: $SU(4)_L\times U(1)_X\to SU(3)_L\times U(1)_{X'}\to SU(2)\times U(1)_Y \to U(1)_{em}$, i.e, the 3-4-1 model can be considered as the extended version of 3-3-1 models \cite{CarcamoHernandez:2022fvl, Oliveira:2022vjo}. The electric operator is defined as follows:
\begin{equation}\label{eq_Qoperator}
	Q= T_3 + \frac{1}{\sqrt{3}} T_8 -\frac{2}{\sqrt{6}} T_{15} + X \mathbb{I},
\end{equation}
where $T_{3,8,15}$ are the diagonal generators of the $SU(4)$ group and $X$ is the $U(1)$ charge, see precise forms  corresponding to the quadruplet  in Ref. \cite{Georgi:2000vve}, for example.

The lepton sector  consists of three left-handed quadruplets and respective right-handed singlets, namely 
\begin{align}
	\label{eq_lepton}	
&	L_{a}   =   ( \nu'_a\, , e'_a \, , E'_a\, ,N'_a )_L^T \sim \left(1, 4,-\frac{1}{2} \right)\, ,\; a=1,2,3, 
\crn &	e'_{a R},\; E'_{a R}  \sim   (1, 1, -1), \hs \nu'_{aR},\;N'_{a R} \sim (1, 1,0). 
\end{align}
The Higgs multiplets  and non-zero vacuum expectation values (VEV) of neutral components  needed for generating all fermion masses are: 
\begin{align}
	\label{eq_Higgsmultiplet}
	\chi& = \left(%
	\chi_1^{0} \, ,
	\chi_2^{-} \, ,
	\chi_3^{-} \, ,
	\chi_4^{0}
	\right)^T \sim \left( 1, 4, -\frac{1}{2}\right), \; \langle \chi \rangle =    \left(%
	0 \, ,
	0\, ,
	0 \, ,
	\fr{V}{\sqrt{2}}
	\right)^T, 
	\crn  \phi &= \left(%
	\phi_1^{+}  \, ,
	\phi_2^{0} \, ,
	\phi_3^0 \, ,
	\phi_4^{+}
	\right)^T\, \sim \left( 1, 4,\frac{1}{2}\right), \; \langle \phi \rangle  =    \left(%
	0 \, ,
	0\, ,
	\fr{\om}{\sqrt{2}}\, ,
	0
	\right)^T, 
	\crn \rho& = \left(%
	\rho_1^{+}  \, ,
	\rho_2^{0} \, ,
	\rho_3^{0} \, ,
	\rho_4^{+}
	\right)^T \sim \left( 1, 4, \frac{1}{2}\right), \; \langle \rho  \rangle = \left(%
	0 \, ,
	\fr{v_1}{\sqrt{2}}\, ,
	0 \, ,
	0
	\right)^T, 
	\crn \eta &= \left(%
	\eta_1^{0}  \, ,
	\eta_2^- \, ,
	\eta_3^{-} \, ,
	\eta_4^{0}
	\right)^T\, \sim \left( 1, 4, -\frac{1}{2}\right), \;  \langle \eta \rangle  =    \left(%
	\fr{v_2}{\sqrt{2}} \, ,
	0\, ,
	0\, ,
	0
	\right)^T. 
\end{align}
The lepton masses  are generated from the following Yukawa  interactions
\begin{align}
	\label{eq_Lyfermion}
	-\mathcal{L}_y= & \;  Y^{N}_{a b} \overline{L_{a} } \chi  N'_{b R}  + Y^{E}_{a b} \overline{L_{a}} \phi E'_{b R}  + Y^{e}_{a b} \overline{L_{a} } \rho e'_{b R} +  Y^{\nu}_{a b} \overline{L_{a} } \eta  \nu'_{b R}
	+\mathrm{ H. c.}. 
\end{align}
The model consists of quark multiplets  that must be arranged to cancel the gauge anomalies, see for example a discussion in Ref. \cite{Long:2016lmj}. It can be seen that the quark masses can be constructed to satisfy the recent experimental data.

The zero VEV values of some neutral Higgs components  can be explained by considering a global symmetry called the general lepton number ${\cal L}$ defined as follows: 
\be L =  \fr{4}{\sqrt{3}} \left( T_8 + \fr{1}{\sqrt{2}} T_{15}\right) + {\cal L}\,,
\label{lm3}
\ee
where $L$ is the normal lepton number.  This formula is an  extension of that introduced for a 3-3-1 model \cite{Chang:2006aa}.  As a consequence, all singlets have $L(\mathrm{singlet})={\cal L}(\mathrm{singlet)}$, namely  $\mathcal{L}(u_{a R})=\mathcal{L}(d_{aR})=0$,  $\mathcal{L}(e_{a R})=\mathcal{L}(\nu'_{aR})=1= -\mathcal{L}(E_{a R})=-\mathcal{L}(N'_{aR})$.  The normal lepton  $L$  for a quadruplet   is  computed as follows 
\be
L = \textrm{diag}
\left(%
1 + {\cal L}\, ,  1 + {\cal L} \, , - 1 + {\cal L} \, , -1 + {\cal L}
\right).
\label{m4eq62}
\ee
In this work, we adopt only the Yukawa couplings respecting  the generalized lepton  number $\mathcal{L}$ including the Lagrangian \eqref{eq_Lyfermion} for leptons. The particular values of  $\mathcal{L}$ of all multiplets are listed in Tables \ref{bcharge}.
\begin{table}[ht]
	\centering  
	\newcolumntype{C}{>{\centering\arraybackslash}p{0.4cm}}
\renewcommand*{\arraystretch}{1.2}%
\begin{tabular}{l|CCCC|CCCCc|CCCCCCCC}
	\hline 
	Multiplet & $\chi$ &$\ph$ & $\eta$ & $\rho$  &  $L_{a}$& $\nu'_{aR}$ & $e'_{aR}$&$E'_{aR}$& $N'_{aR}$ &$Q_{\al L}$ & $Q_{3L}$ &
	$u'_{aR}$&$d'_{aR}$ &$D'_{\al R}$ &$U'_{\al R}$ & $U'_{3R}$ &$D'_{3 R}$  \\
	\hline $\cal L$ charge &$1$ & $1$ &$-1 $ &  $- 1 $  &
	$ 0  $&1 &1&-1&-1 &
	$ 1  $ & $ -1 $&$ 0$ & $0$ & $2$& $2$& $-2$&$-2$\\
	\hline 
\end{tabular}
	\caption{ ${\cal L}$ charges for multiplets in
		the 341RH. }	\label{bcharge}
\end{table}
 They result in consistent values of the normal lepton numbers for  all SM leptons $L(\nu'_{aL,R})=L(e'_{aL,R}) =1$, and  all SM quarks have $L=0$. The  remaining non-zero lepton numbers $L$  are listed in Table \ref{table_lnumber},
\begin{table}[ht]
	\centering 
	{
	\newcolumntype{C}{>{\centering\arraybackslash}p{0.75cm}}
	\newcolumntype{D}{>{\centering\arraybackslash}p{0.35cm}}
	\renewcommand*{\arraystretch}{1.2}%
	\begin{tabular}{l|DD|DD|DD|DD|CCCCCC}
		\hline
		Fields
		&$\chi_1^0$&$\chi_2^-$&$\phi^+_1$
		& $\ph_2^{0}$&$\rho^{0}_3$&$\rho_4^+$&$ \eta_3^-$&$ \eta_4^0 $ & $E'_{aL,R}$ & $N'_{aL,R}$  & $U'_{\alpha L,R}$ & $D'_{\alpha L,R}$& $U'_{3 L,R}$ & $D'_{3 L,R}$ \\
		\hline
		$L$ & $2$ & $2$ & $2$ & $2$
		&$-2$&$-2$&$-2$&$-2$&-1&-1&2&2 &-2&-2\\
		\hline
	\end{tabular}
}
	\caption{
	Nonzero lepton number $L$ of all fields  in the 341RH.} 
	\label{table_lnumber}
\end{table}
because of the requirement that the total lepton number $L$ is always conservative. 

The mass terms of all leptons are:
\begin{align} \label{eq_chargedmass}
	(M_{\nu})_{a b}= Y^\nu_{a b} \fr{v_2}{\sqrt{2}}\,,\; (M_{e})_{a b}= Y^e_{a b} \fr{v_1}{\sqrt{2}}\,, \; (M_{E })_{a b} = Y^E_{a b} \fr{\om}{\sqrt{2}}, \; (M_{N })_{a b} = Y^N_{a b} \fr{V}{\sqrt{2}}.
\end{align}
  The active Dirac neutrino masses and mixing are constructed  from the mass matrix $M_{\nu}$.  But, these tiny masses do not affect significantly the one-loop contributions to $a_{e_a}$. 

Now we focus on the lepton sector in the Yukawa part of Eq. \eqref{eq_Lyfermion}.  The normal lepton mass matrix $M_e$ given in Eq.~\eqref{eq_chargedmass} is assumed to be diagonal for simplicity. As a result, the flavor basis of the charged leptons $e'_a$ is the mass basis $e_{aL,R}\equiv e'_{aL,R}$, namely 
\begin{equation}\label{eq_SMmasse}
	m_{e_a}= Y^e_{ab}\delta_{ab} \frac{v_1}{\sqrt{2}} \Rightarrow Y^e_{ab}=\delta_{ab}\frac{\sqrt{2}m_{e_a}}{v_1}. 
\end{equation}
Three other base $f'_{L,R}\equiv (f'_1, f'_2, f'_3)^T_{L,R}$ with $f=\nu,E,N$  are transformed  into the corresponding mass base $f'_{L,R}$ through the following relations:
\begin{align}
	\label{eq_Lmixing}
	U^{f\dagger}_L M_{\nu}U^{f}_R &=\hat{M}_{f}=\mathrm{diag}(m_{f_1},m_{f_2},m_{f_3}),\; f'_{L,R} =U^{f}_{L,R}f_{L,R}. 
\end{align}


Although the quark sector is irrelevant to our work, we review here the main property relating to the recent experimental constraints of $K^0-\bar{K}^0$ oscillation, similar to the 3-3-1 models \cite{CarcamoHernandez:2022fvl, Oliveira:2022vjo}. The quark sector consists of two families of left-handed anti-quadruplets, one family of left-handed quadruplet, and singlets of right-handed quarks, namely
\begin{align}
	Q_{3 L} &  =  \left(%
	u'_3  ,
	d'_3 ,
	D'_3,
	U'_3
	\right)^T_L \sim \left(3, 4,    \fr{1}{6} \right), \; Q_{\al L}   = \left(%
	d_\al  ,
	- u_\al  ,
	U'_\al ,
	D'_\al
	\right)^T_L \sim \left(3, 4^*,  \frac{1}{6}\right) , \; \al = 1, 2 , \crn
	u'_{iR}  &\sim  (3, 1, 2/3 ),\;  d'_{i R}  \sim  (3, 1, - 1/3 ), \;  
	D'_{iR}  \sim   \left(3, 1, -1/3\right) ,\; U^\prime _{iR} \sim \left(3, 1, 2/3\right),\; i=\overline{1,3}.
	\label{eq_Q3}
\end{align}
The relevant  Yukawa parts of quarks respecting all gauge and $\mathcal{L}$ symmetries  are
\begin{align}
	- \mathcal{L}_{Y}^{q}  =&   \sum_{i=1}^3\left[ \sum_{\alpha=1}^2\left( Y^{u}_{\al i}\overline{Q_{\al  L} } \rho^* u_{i R} + Y^{d}_{\al i}\overline{Q_{\al  L} } \eta^* d_{i R}\right) + \left( Y^{u}_{3 i}\overline{Q_{3  L} } \eta u_{i R} + Y^{d}_{3 i}\overline{Q_{3  L} } \rho d_{i R}\right)\right]     
	\crn&+  \sum_{\alpha,\beta=1}^2  \left[Y^{U}_{\al \bet}\overline{Q_{\al  L} } \phi^*U_{\bet R}  +  Y^{D}_{\al \bet}\overline{Q_{\al  L} } \chi^* D^\prime_{\bet R}\right]    + Y^{U}_{33}\overline{Q_{3  L} } \chi U_{\bet R}  + 
	Y^{D}_{33}\overline{Q_{3  L} } \eta D^\prime_{\bet R}
	\crn & +  \mathrm{H. c.},
	\label{eq_Lyq}
\end{align}
We can see that the Yukawa parts generating SM quark masses are the same as those shown in Refs. \cite{CarcamoHernandez:2022fvl, Oliveira:2022vjo} for 3-3-1 models with right-handed neutrinos, where both $\rho$ and $\chi$ inherit the same VEV properties. In addition, all exotic quarks do  not mix which the SM quarks, consistent with the property shown in Table \ref{table_lnumber} that all SM quarks have zero $L$, in contrast with $L=\pm2$ for all exotic quarks. Therefore, the couplings of the SM quarks with Higgs bosons in the models given in Refs. \cite{CarcamoHernandez:2022fvl, Oliveira:2022vjo}  are the same as those presented in the model under consideration. Accordingly,  the mass split value $\Delta m_K$ originated from the $K^0$-$\bar{K}^0$ oscillation predicted by the 3-4-1 model under considerations will be the sum of the new heavy  Higgs and  neutral gauge contributions at the tree level, in which every contribution is proportional to the inverse of the squared mass of the respective Higgs or gauge boson. As we will see below, the 3-4-1 model predicts two new heavy neutral gauge bosons $Z_3$ and $Z_4$ with masses  $m^2_{Z_3}\varpropto w^2$ and $m^2_{Z_4} \varpropto V^2$ in the limit $V\gg w$. Because $w$ plays the role of the $SU(3)_L$ scale, Ref. \cite{CarcamoHernandez:2022fvl} confirms that if $m_{Z_4}\gg m_{Z_3}>4 $ TeV, contributions from these two gauge bosons to  $\Delta m_K$ will satisfy the experimental constraint. The case of the heavy Higgs contributions are the same. Because this choice of parameters does not affect the main results in our work, this experimental constraint will be ignored in the remaining part of this work.

\subsection{\label{gaugeboson} Gauge boson masses and mixing}
Gauge boson masses arise from the covariant kinetic term of Higgs multiplets, namely 
\begin{equation}\label{eq_LkH}
	L_{\mathrm{Higgs}} = \sum_{H}^4 \left(D^\mu\langle H \rangle\right)^\dag D_\mu \langle H \rangle ,
\end{equation}
where $H=\chi, \phi,\eta,\rho$. The covariant derivative is defined as
\begin{align} \label{eq_defDmu}
	D_\mu & = \pa_\mu - i g\sum_{a=1}^{15} W_{ a\mu} T_a  - i g_X X B''_\mu T_{16}
	\equiv  \pa_\mu - i g P_\mu^{NC} - i g P_\mu^{CC}\, ,
\end{align}
where $ g, g_X$ and $ W_{a \mu} , B''_\mu$ are gauge couplings  and fields of  the gauge groups $SU(4)_L$ and $U(1)_X$, respectively.  The two parts $P_\mu^{NC}$ and $P_\mu^{CC}$ relate to the  neutral and non-hermitian currents \cite{Long:2016lmj}.
For quadruplet, $T_{16} = \fr{1}{2\sqrt{2}} \textrm{diag} (1,1,1,1)$ and 
\begin{align}
	\label{eq_defPCC}
	P_\mu^{CC} =& \fr{1}{\sqrt{2}}\left(%
	\begin{array}{cccc}
		0 & W^+ &W_{13}^{+} & W_{14}^{0}\\
		W^-  & 0 & W_{23}^{0}  &W_{24}^{-}\\
		W_{13}^{-} & W_{23}^{0*} & 0 &W_{34}^{-} \\
		W_{14}^{0*}&W_{24}^{+}&W_{34}^{+}&0
	\end{array}\,
	\right)_\mu\, ,
\end{align}
where  $t\equiv g_X/g$  and  $\sqrt{2}\, W^\mu_{ij} \equiv W^\mu_i - i W^\mu_j$ with $i<j$.
The upper subscripts label the electric charges of gauge bosons. The relation between the original basis $(W_3,\; W_8,W_{15}, B'')$ and the mass basis $(A,Z,Z_3,Z_4)$  of all real neutral boson was determined previously \cite{Long:2016lmj}. 
  The  masses of non-Hermitian (charged)  gauge bosons are given by
\begin{align}
	m^2_W &  =  \fr{g^2(v_1^2 + v_2^2)}{4}\, ,  m^2_{W_{13}}  =  \fr{g^2(v_2^2 + \om^2)}{4}\, ,
	m^2_{W_{23}}  =  \fr{g^2(v_1^2 + \om^2)}{4}\, , \crn
	m^2_{W_{14}} & =     \fr{g^2(v_2^2 + V^2)}{4}\, , m^2_{W_{24}}  =  \fr{g^2(v_1^2 + V^2)}{4}\, ,
	m^2_{W_{34}}  =  \fr{ g^2(\om^2 + V^2)}{4}\, .
	\label{eq_mWpm}	
\end{align}
By  spontaneous symmetry breaking (SSB), the following relation should be in order:
$V\gg \om \gg v_1, v_2$.  A consequence from Eq. \eqref{eq_mWpm} is  that $W^\pm$ must be identified with the singly charged  SM gauge boson, namely 
\be v_1^2 + v_2^2 = v^2= 246^2\; \textrm{GeV}^2. \ee
Then neutral gauge boson masses are 
\begin{align}
	\label{eq_mV0}	
	m^2_{Z} &\simeq \frac{m^2_W}{c^2_W}, \; m^2_{Z_3} \simeq \frac{4g^2 c_W^2 w^2}{3(1-4s_W^2)},\; m^2_{Z_4} \simeq \frac{g^2 (9V^2 +w^2)}{6}.
\end{align}
The above formula implies that $\eta$ and $\rho$ play roles as two Higgs doublets in the well-known two Higgs doublet models  after the breaking steps to the SM gauge group  $SU(2)_L\times U(1)_Y$.  Then we define the mixing angle $\beta$  as follows
\begin{equation}\label{eq_tb}
	t_{\beta}\equiv \tan\beta= \frac{v_2}{v_1},\; v_1=v c_{\beta}, \; v_2=vs_{\beta},
\end{equation}
where $t_{\beta}\geq0.4$ as the perturbative limit of the top quark Yukawa coupling $|Y^u_{33}| \simeq \sqrt{2} m_t/(v s_{\beta}) <\sqrt{4\pi}$. The upper bound of $t_{\beta}$ may comes from the  tau mass $Y^{\tau}\simeq \sqrt{2}m_{\tau}/(vc_{\beta})<\sqrt{4\pi}$, equivalently $t_{\beta}<390$.

The mixing parameters of the neutral gauge bosons were summarized in appendix \ref{app_Zmm}, see the details in Ref. \cite{Long:2016lmj}.  Because the current study \cite{CarcamoHernandez:2022fvl}  shows that the breaking scale of $SU(3)_L$ and $SU(L)_L$ is the order of TeV, the mixing parameters between $Z$, $Z_3$, and $Z_4$ are much small, therefore we ignore in this work.  But the loop corrections to the $Z\mu^+\mu^-$ may be significant, especially in the model inheriting the "chiral enhancement" enough to accommodate the $(g-2)_{e_a}$ anomaly data. These one-loop corrections were computed in appendix \ref{app_Zmm} used to constrain the experimental data in the numerical investigation.

\subsection{\label{potential}Higgs potential and Higgs spectrum}
The most general Higgs potential including the appearance of the singly charged Higgs boson $\sigma^\pm\sim (1,1,\pm1)$ is:
\begin{align} \label{eq_Vh}
	V_h =&\; V(\eta,\rho,\ph,\chi)+ V(\sigma^\pm),
	\crn  	V(\eta,\rho,\ph,\chi) =&\; \mu^2_1 \eta^\dag \eta +
	\mu^2_2  \rho^\dag \rho +  \mu^2_3  \ph^\dag \ph + \mu^2_4  \chi^\dag \chi \crn
	&+\lambda_1 (\eta^\dag \eta)^2 + \lambda_2(\rho^\dag \rho)^2 +
	\lambda_3  (\ph^\dag \ph)^2  + \la_4 ( \chi^\dag \chi)^2 \crn
	& +  (\eta^\dag \eta) [ \lambda_5 (\rho^\dag \rho) +
	\lambda_6 (\ph^\dag \ph)  + \lambda_7 (\chi^\dag \chi) ]\crn
	&+ \ (\rho^\dag \rho)[ \lambda_8(\ph^\dag \ph) + \la_9 (\chi^\dag \chi)] +\lambda'_9 (\ph^\dag \ph) (\chi^\dag \chi) \crn
	& +  \lambda_{10} (\rho^\dag \eta)(\eta^\dag \rho) +
	\lambda_{11} (\rho^\dag \ph)(\ph^\dag \rho) + \lambda_{12} (\rho^\dag \chi)(\chi^\dag \rho) \crn
	& +   \lambda_{13} (\ph^\dag \eta)( \eta^\dag \ph) +\la_{14}\, (\chi^\dag \eta)( \eta^\dag \chi)
	+\la_{15}\, (\chi^\dag \ph)( \ph^\dag \chi) \crn
	&+ (f \ep^{i j k l} \eta_i \rho_j \ph_k \chi_l +\mathrm{ H.c.}),
	\\ V(\sigma^\pm) =&\; \mu_{5}^2 \sigma^+\sigma^- + \lambda_{\sigma} \left( \sigma^+\sigma^-\right)^2 + \left( f_{\sigma}\sigma^+  \rho^\dagger\eta +\mathrm{ H.c.}\right)
	\crn &+ (\sigma^+\sigma^-)  \left( \lambda_{\sigma \eta} \eta^\dag \eta + 
	\lambda_{\sigma \rho}  \rho^\dag \rho + \lambda_{\sigma \phi}  \ph^\dag \ph + \lambda_{\sigma \chi}  \chi^\dag \chi\right). %
	\label{eq_VSigmapm}
\end{align}
Here we consider the Higgs potential respecting the generalized  lepton number $\mathcal{L}$ and the new singly charged Higgs boson has $\mathcal{L}(\sigma^\pm)=0$.  We also ignore the triple Higgs self couplings $(f'_{\sigma} \chi^\dagger \phi +\mathrm{h.c.})$ because it results in unnecessary mixing  in our calculation. The detailed discussion to derive the masses and mixing parameters of the Higgs bosons were presented previously  in Ref. \cite{Long:2016lmj} without  $\sigma^\pm$. The Higgs potential consisting of $\sigma^\pm$  was  discussed in 3-3-1 models \cite{Hue:2021zyw, Hue:2021xzl}, in which the detailed calculation to derive Higgs masses and mixing were performed. We collect only the most important results relating to our work.

The mixing of $\chi_2^{\pm}$ and $\rho^\pm_4$ results in two massless Goldstone boson $G^{\pm}_{24}$ and  two singly charged Higgs bosons $h^{\pm}_3$:
\begin{align}
	\label{eq_chi2rho4}
	\begin{pmatrix}
		\chi_2^{\pm}	\\
		\rho_4^{\pm}	
	\end{pmatrix}
	=\left(
	\begin{array}{cc}
		c_{\theta_1} & s_{\theta_1} \\
		-s_{\theta_1} & c_{\theta_1} \\
	\end{array}
	\right)	\begin{pmatrix}
		G_{24}^{\pm}	\\
		h_3^{\pm}	
	\end{pmatrix},\;  M^2_{h^+_3}= \left(c_{\beta }{}^2 v^2+V^2\right) \left(\frac{\lambda _{12}}{2}-\frac{f t_{\beta } w}{2  V}\right), 
\end{align}
and 
\begin{equation}\label{eq_theta1}
	s_{\theta1}\equiv \sin\theta_1,\; 	c_{\theta1}\equiv \cos\theta_1,\;  \tan\theta_1=\frac{vc_{\beta}}{V}. 
\end{equation}

We consider  here the mixing of $\phi^{0*}_2$ and $\rho^0_3$, which leads to a non-hermitian Goldstone boson $G^0_{23}\neq G^{0*}_{23}$ and a physical state $H^0_1\neq H^{0*}_{1}$
\begin{align}
	\label{eq_phi2rho3}
	\begin{pmatrix}
		\phi_2^{0*}	\\
		\rho_3^{0}	
	\end{pmatrix}
	=\left(
	\begin{array}{cc}
		c_{\theta_2} & s_{\theta_2} \\
		-s_{\theta_2} & c_{\theta_2} \\
	\end{array}
	\right)	\begin{pmatrix}
		G_{23}^{0}	\\
		H_1^{0}	
	\end{pmatrix},\;  M^2_{H^0_1}= \left(c_{\beta }{}^2 v^2+w^2\right) \left(\frac{\lambda _{11}}{2}-\frac{f t_{\beta } V}{2 w}\right), 
\end{align}
and 
\begin{equation}\label{eq_theta2}
	s_{\theta_2}\equiv \sin\theta_2,\; 	c_{\theta_2}\equiv \cos\theta_2,\;  \tan\theta_2=\frac{vc_{\beta}}{w}. 
\end{equation}
Three  singly charged Higgs bosons $( \rho^\pm_1,\; \eta^\pm_2,\sigma^\pm)$ are  changed into  the two  physical states $h^\pm_{1,2}$ and the Goldstone bosons $G^\pm_W$ of $W^\pm$ as follows:
\begin{align}
	\rho^{\pm}_1&= c_{\beta} \phi^\pm_W +s_{\beta} \left( c_{\alpha} h^\pm_1 +s_{\alpha} h^{\pm}_2\right), 
	\crn 	\eta^{\pm}_2&= -s_{\beta} \phi^\pm_W + c_{\beta} \left( c_{\alpha} h^\pm_1 +s_{\alpha} h^{\pm}_2\right), 
	\crn  	\sigma^{\pm}&=   -s_{\alpha} h^\pm_1 +c_{\alpha} h^{\pm}_2. 
	\label{eq_scHigg}
\end{align}
The relations in Eq. \eqref{eq_scHigg} were given in Refs. \cite{Hue:2021zyw, Hue:2021xzl}, which  consist of  the same part $V(\sigma)$ given in Eq. \eqref{eq_VSigmapm} in the  Higgs potential.  The mixing parameter $\alpha$ and Higgs boson masses $m_{h^+_1}$, $m_{h^+_2}$ will be investigated as free parameters, while  three dependent parameters are:
\begin{align}
	\label{eq_Higgscouplings}
	\mu _5^2&=  c_{\alpha }{}^2 m_{h_2^+}^2 + 
	s_{\alpha }{}^2 m_{h_1^+}^2 - \frac{1}{2} \left(c_{\beta }{}^2 \lambda_{\sigma \rho } v^2 +\lambda _{\sigma \chi } V^2 +\lambda_{\sigma \phi } w^2  +\lambda _{\sigma \eta } s_{\beta }{}^2 v^2\right), 
	\crn f &=-\frac{c_{\beta } s_{\beta } \left(2 c_{\alpha }{}^2 m_{h_1^+}^2-\lambda _{10}
		v^2+2 s_{\alpha }{}^2 m_{h_2^+}^2\right)}{V w},
	\crn f_{\sigma }&= -\frac{\sqrt{2} c_{\alpha } s_{\alpha } (m_{h_1^+}^2-m_{h_2^+}^2)}{v}. 
\end{align} 

As usual, this model must contain a Standard model-like (SM-like) Higgs boson confirmed by LHC. A detailed calculation to identify this Higgs boson in the model under consideration can be found in Ref. \cite{Long:2016lmj}.   A simple estimation is presented in appendix \ref{app_SMlikeHiggs}. In general, the SM-like Higgs boson gets main contributions from $\eta$ and $\rho$, similarly to the case of the property of the well-known two Higgs doublet model (2HDM). In the model under consideration, the flavor and the physical base of charged leptons are the same, as assumed in Eq. \eqref{eq_SMmasse}.  As a result, there is no mixing between SM and heavy charged leptons, implying that the leading corrections to the SM coupling  $h\mu^+\mu^-$ is the combination of the tree-level of  the Higgs mixing and the one-loop level \cite{Dermisek:2013gta, Crivellin:2021rbq, Escribano:2021css, Crivellin:2022wzw}. As commented in Ref. \cite{Escribano:2021css}, the recent experiments give no hint for $h\to e^+ e^-$, while only the evidence of   $h\to \mu^+ \mu^-$ was reported \cite{CMS:2020xwi, ATLAS:2020fzp}.  The loop corrections to the coupling  $h\mu^+\mu^-$  is smaller than the recent experimental sensitivities, but may be detected in the future \cite{Crivellin:2021rbq, Escribano:2021css, Crivellin:2022wzw}. Therefore, the constraint from the modification of the coupling $h\mu^+\mu^-$ will not affect the allowed regions of the parameter space we will discuss in this work.

\subsection{Analytic formulas for one-loop contributions to $a_{e_a}$ with active Dirac neutrinos}
In this section, we consider the simple case that $\nu_{1,2,3}$  are Dirac ones, and no mixing between $\sigma^\pm$ and  other singly charged Higgs bosons, i.e. $s_{\alpha}=0$, $c_{\alpha} =1 $, then only $h_2^\pm$ couples with active neutrinos. The conclusion with $s_{\alpha}\neq 0$ is unchanged because the tiny neutrino masses result in suppressed one-loop contributions to $(g-2)_{e_a}$ anomalies. 

The relevant Lagrangian giving one-loop contributions to $a_{e_a}$ is:
\begin{align}
	\label{eq_Lylepton1}
	-\mathcal{L}^l_y &  = \frac{g}{\sqrt{2} m_{W_{23}}} \sum_{a,c=1}^3 U_{L,ac}^{E*} \left[ m_{E_c}t_{\theta_2}P_L +m_{e_a}t^{-1}_{\theta_2}P_R  \right]e_aH_1^0 \crn
	&+  \frac{g}{\sqrt{2} m_{W_{24}}} \sum_{a,c=1}^3 U_{L,ac}^{N*} \left[ m_{N_c}t_{\theta_1}P_L + m_{e_a}t^{-1}_{\theta_1}P_R \right]e_ah_3^+ 
	\crn&+  \frac{g}{\sqrt{2} m_{W}}  \sum_{a,c=1}^3 \overline{\nu_{c }}  U^{\nu*}_{L,ac}  \left(  m_{\nu_c}t^{-1}_{\beta}P_L + m_{e_a}t_{\beta} P_R \right) e_{a}h^{+}_1 +\mathrm{H.c.} +\dots. 
\end{align}
To avoid large cLFV decays $e_b\to e_a \gamma$, which may be ruled out by experiments, we will pay attention to the limit that $U^N_{L}=U^E_L=I_3$, and $m_{E_{a}}=m_E$, $m_{N_{a}}=m_{N}$ for all $a=1,2,3$. For active neutrinos having tiny masses, the respective one-loop contributions to $\Delta a_{e_a}$ are suppressed. The non-zero form factors relevant with one-loop contributions to $a_{e_a}$ are \cite{Crivellin:2018qmi}
\begin{align}
	a_{e_a}(h_3^+)&=- \frac{g^2m^2_{e_a}}{8\pi^2 m^2_{W}} \left\{ \frac{v^2}{V^2}  \times x_{N} f_{\Phi}(x_{N})  + \left[\frac{v^4c^2_{\beta}}{V^4}  x_{N} +  \frac{ m^2_{e_a}c^{2}_{\theta_1} }{m^2_{h^\pm_3}c^2_{\beta}}\right]\tilde{f}_{\Phi}(x_{N})   \right\},
	\crn a_{e_a}(h_1^+)  &= -\frac{g^2m^2_{e_a}}{8\pi^2m^2_W}  \left( \frac{m_{e_a}t_{\beta}}{m_{h^+_1}}\right)^2 \tilde{f}_{\Phi}(0) ,  \label{eq_caaH}
	\\ a_{e_a}(H_1^0)&= - \frac{g^2m^2_{e_a}}{8\pi^2 m^2_{W}} \left\{ \frac{v^2}{w^2} x_{E} \left[ f_{\Phi}(x_{E}) - g_{\Phi}(x_{E})\right]  
	%
	+ \left(\frac{v^4s^2_{\beta}}{w^4} x_{E} + \frac{ m^2_{e_a}c^{2}_{\theta_2} }{m^2_{H^0_1}c^2_{\beta}} \right) \left[ \tilde{f}_{\Phi}(x_{E}) -\tilde{g}_{\Phi}(x_{E})\right]   \right\}, \nn 
\end{align}
where $ x_{N}=m^2_{N}/m^2_{h^+_3}$   and $ x_{E}=m^2_{E}/m^2_{H^0_1}$, and 
\begin{align}
	\label{eq_fphiX}	
	f_\Phi (x)&= 2\tilde{g}_\Phi(x)=\frac{x^2-1 -2x\ln x}{4(x-1)^3},\crn 
	g_\Phi&=\frac{x-1 -\ln x}{2(x-1)^2}, \crn 
	\tilde{f}_\Phi(x)&= \frac{2x^3 +3x^2 -6x +1 -6x^2 \ln x}{24(x-1)^4}. 
\end{align}

The total contribution  to $\Delta a_{e_a}$ from all Higgs bosons are:
\begin{align}\label{eq_aeaX}
	a_{e_a}(H)&=\sum_X a_{e_a}(X),
	%
\end{align}
where $X=h^+_1,h^+_3,H^0_1$.

\begin{figure}[ht]
	\centering
	\includegraphics[width=9cm]{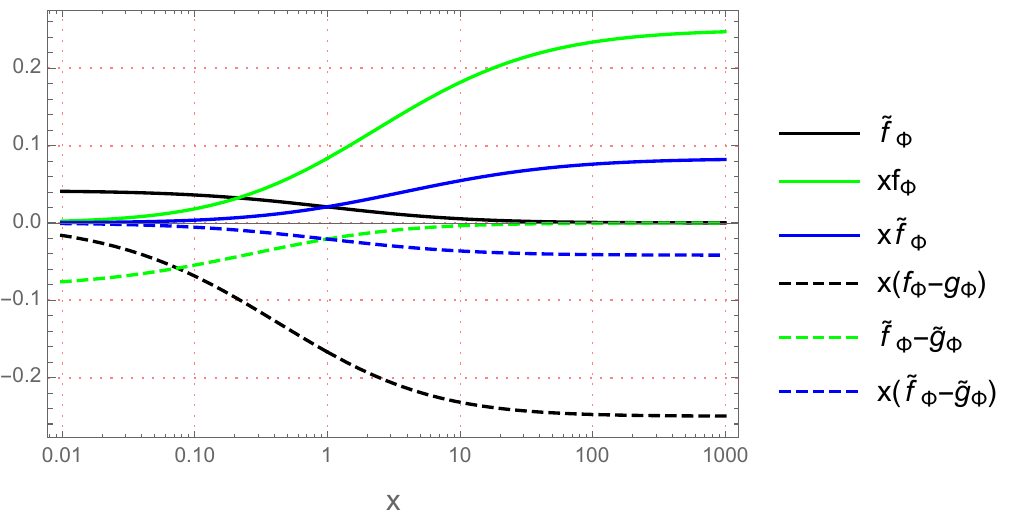}
	\caption{ Properties of the scalar functions relating to one-loop contributions of heavy Higgs bosons to $(g-2)_{e_a}$ anomalies in the 3-4-1 model with active Dirac neutrinos.} \label{fig_fgX}
\end{figure}

The functions relating to one-loop contributions of Higgs bosons are shown in Fig. \ref{fig_fgX}.
Because $xf_{\Phi}(x)$, $\tilde{f}_{\Phi}(x)$,   $x\tilde{f}_{\Phi}(x)>0$ for all $x>0$, the one-loop contributions from singly charged Higgs bosons $h^{\pm}_{1,3}$ are in opposite signs with $\Delta a^{\mathrm{NP}}_{\mu}$, hence they should be small. On the other hand, the one-loop contribution from the neutral Higgs boson $H^0_1$ consists of  negative functions $\left( f_{\Phi}(x)-g_{\Phi}(x)\right)$,  and $\left( \tilde{f}_{\Phi}(x) -\tilde{g}_{\Phi}(x)\right)$, hence the final contributions support large $\Delta a_{\mu}$.

The couplings of  leptons and non-hermitian gauge bosons  are 
\begin{align}
	\mathcal{L}^{\ell\ell V}=& i \overline{L_{aL}}\gamma^{\mu}D_{\mu}L_{aL}
	\crn=&\frac{g}{\sqrt{2}} \left[ U^{\nu*}_{ai} \overline{ \nu_{i}}\gamma^\mu P_L e_{a}W^{+}_{\mu} + U^{N*}_{ai}\overline{N_{i}} \gamma^\mu  P_L e_{a}W^{+}_{24 \mu} +\overline{e_{aL}}\gamma^\mu  E'_{aL} W^{0}_{23 \mu} +\dots,
	\right] +\mathrm{H.c.},\label{llv1}
\end{align}
where the last line collects only terms giving one-loop contributions to cLFV amplitudes and $(g-2)$ anomalies,  resulting in the  following  formulas \cite{Crivellin:2018qmi}, 
\begin{align}
	a_{e_a}(W)&= - \frac{g^2 m^2_{e_a}}{8\pi^2m_W^2} \sum_{i=1}^{3}U_{ai}^\nu U_{ai}^{\nu*} {\tilde{f}_{V }\left({x_{\nu_i} }\right)} \simeq - \frac{g^2 m^2_{e_a}}{8\pi^2m_W^2}{\tilde{f}_{V}\left(0\right)} ,  \label{eq_aeaWDirac}
	\\ a_{e_a}(W_{24}) &=  - \frac{g^2 m^2_{e_a}}{8\pi^2m_W^2} \sum_{b=1}^{3}U_{ab}^N U_{ab}^{N*}\frac{m_W^2}{m^2_{W_{24}}} {\tilde{f}_{V}\left({x_{N_b} }\right)} =- \frac{g^2 m^2_{e_a}}{8\pi^2m_W^2}\times \frac{m_W^2}{m^2_{W_{24}}} {\tilde{f}_{V}\left({x_{N} }\right)}, \label{eq_aeaW24}
	\\ a_{e_a}(W_{23}) &=  - \frac{g^2 m^2_{e_a}}{8\pi^2m_W^2} \sum_{b=1}^{3}U_{ab}^E U_{ab}^{E*}\frac{m_W^2}{m^2_{W_{23}}} {\tilde{f}_{V}\left({x_{E_b} }\right)} =- \frac{g^2 m^2_{e_a}}{8\pi^2m_W^2}\times \frac{m_W^2}{m^2_{W_{23}}} {\tilde{f}_{V}\left({x_{E} }\right)}, \label{eq_aeaW23}
\end{align}
where 
\begin{equation}\label{eq_fVX}
	\tilde{f}_V(x) = \frac{-4x^4 +49x^3 -78 x^2 +43x -10 -18x^3\ln x}{24(x-1)^4}.
\end{equation}
The total one contribution  from new heavy gauge bosons to the $a_{e_a}$ is
\begin{align}
	a_{e_a}(V) =  a_{e_a}(W_{23}) +a_{e_a}(W_{24})  . 
	\label{eq_aeaV}
\end{align}
The deviation of $a_{\mu}$ between predictions by  the two models 3-4-1 and SM are
\begin{align}
	\label{eq_defDamu341}
	\Delta a^{\mathrm{341}}_{e_a}&\equiv \Delta a_{e_a}= \Delta a_{e_a} ({W})+ a_{e_a}(V) + a_{e_a}(H), \quad  \Delta a^{W}_{e_a}= a^{W}_{e_a} - a^{\mathrm{SM}}_{e_a}(W), 
\end{align}
where $a^{\mathrm{SM}}_{\mu}(W)=3.887\times 10^{-9}$~\cite{Jegerlehner:2009ry}, and  $a^{\mathrm{SM}}_{e_a}(W)=a^{\mathrm{SM}}_{\mu}(W)\times (m_{e_a}^2/m^2_{\mu})$ are  the SM's prediction for the one-loop contribution from $W$ boson.  In this work, $\Delta a^{\mathrm{341}}_{e_a}=\Delta a_{e_a}$ will be considered as new physics (NP) predicted by the 3-4-1 models, which must satisfy the  experimental data  given in Eqs. \eqref{eq_damu} and \eqref{eq_dae} in the numerical investigation.

\begin{figure}[ht]
	\centering
	\begin{tabular}{cc}
		\includegraphics[width=9cm]{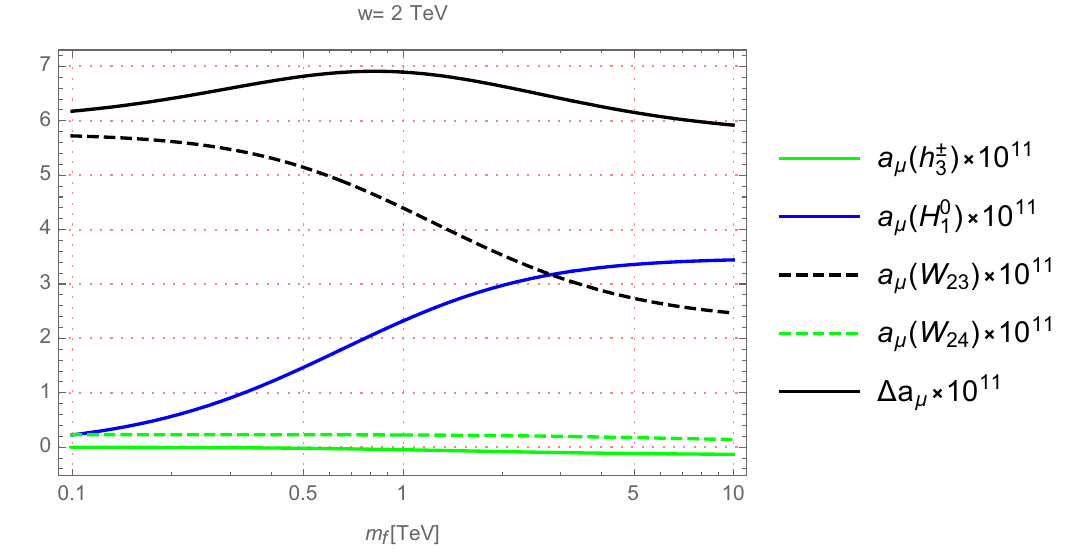}&
		\includegraphics[width=6.5cm]{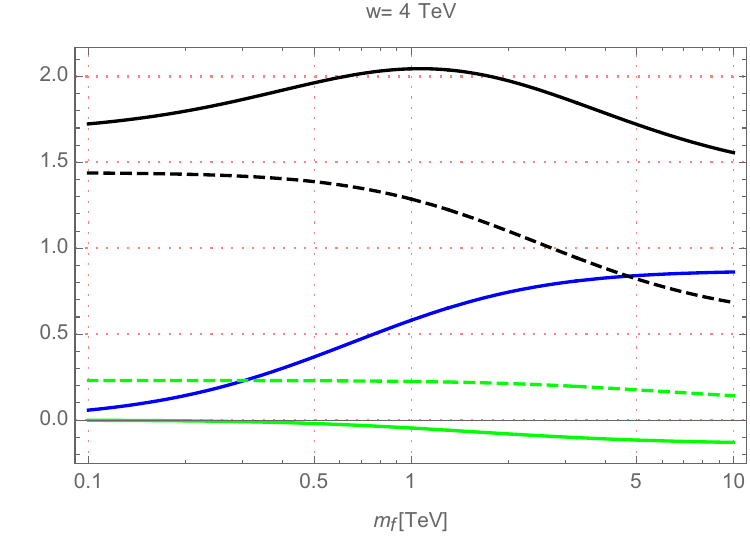}\\	
	\end{tabular}
	\caption{ One-loop contributions of new $SU(4)_L$ particles to $a_{\mu}$ with  $w=2.$ TeV (left panel) and $w=4$ TeV (right panel).} \label{fig_amuH}
\end{figure}

Numerical illustrations are shown in Fig. \ref{fig_fgX} with fixed $w=5$ TeV, $V=10$ TeV, $m_{h^\pm_1}=m_{h^\pm_3}=m_{H^0_1}=1$ TeV, and $t_{\beta}=50$.   Different contributions are considered as functions of  $m_N=m_E=m_f$ with numerical illustrations given in Fig.~\ref{fig_amuH}.
Both $a_{e_a}(W)$ and $a_{e_a}(h^\pm_1)$ are independent with $m_f$, $w$, and $V$.  In addition,  the one-loop contribution from $W^\pm$ gauge boson is $a_{e_a}(W)=a^{\mathrm{SM}}_{e_a}(W)$, hence it does not affect $\Delta a_{\mu}$.  The two conditions $t_{\beta}\le 100$ and $m_{h^\pm_1}\ge 200$ GeV give $0<a_{\mu}(h_1^+) \leq 10^{-12}$ hence $a_{\mu}(h_1^+)$ gives suppressed contributions to $ \Delta a_{\mu}$.  The $\Delta a_{\mu}$ depends strongly on $w$, which lower bound is constrained strictly from the masses, which are given in Eq.  \eqref{eq_mV0}, of heavy neutral gauge bosons $Z_{3,4}$, namely $m_{Z_{3,4}}>3$ TeV from LHC searches \cite{Lee:2014kna}.  As a result, large values of  $w\ge2$ TeV give $\Delta a_{\mu}\le 7\times 10^{-13}\ll 192\times 10^{-13}\leq a^{\mathrm{NP}}_{\mu}$. In conclusion, all the one-loop contributions mentioned above   are much smaller than $a^{\mathrm{NP}}_{\mu}$.

\section{\label{sec_lSSmodel} The 341ISS with ISS neutrinos}

Now we consider an extension  of the above  341RHN model that may explain successfully the $(g-2)_{e_a}$ data. This version consists of six right-handed neutrinos  $\nu_{aR}, X_{aR}\sim (1,0)$, $a=1,2,3$ generating  active neutrino masses through the ISS mechanism, and   a singly charged Higgs boson $\sigma^{+}\sim (1,1,1)$  needed to give large one-loop contributions to AMM. We call this model the 3-4-1 model with ISS neutrinos (341ISS).   The generalized lepton numbers are $\mathcal{L}(\nu_{aR})=1$, $\mathcal{L}(X_{aR})=-1$, and $\mathcal{L}(\sigma^{+})=0$. Now tree-level neutrino masses and mixing angles arise from the ISS  mechanism.  Requiring that $\mathcal{L}$ is only softly broken, the additional Yukawa part is
\begin{align}
	-\mathcal{L}_{Y,\nu_R}=& Y^{\nu}_{ab} \overline{\nu_{ aR}} \eta^{\dagger}L_{b}  + (M_{R})_{ab}\overline{\nu_{aR}}(X_{bR})^c   +\frac{1}{2} (\mu_{X})_{ab}\overline{X_{aR}}(X_{bR})^c 
	\crn&+ Y^{\sigma}_{ab} \overline{(X_{aR})^c} e_{bR} \sigma^+  + \mathrm{h.c.},
	\label{eq_SSterm}
\end{align}
where $Y^{\nu}$, $M_R$, $\mu_X$, and  $Y^{\sigma}$ are  $3\times3$ matrices. The first term in Eq.~\eqref{eq_SSterm} is similar to that given in Eq. \eqref{eq_Lyfermion}, but it  will generate the  Dirac neutrino mass matrix $M_D$ instead of the active neutrino masses. 

Notations for flavor states of active left-handed neutrinos are $ \nu_{L}=(\nu'_{1},\nu'_{2},\nu'_{3})^T_L$ and  $ \nu_R=(\nu_{1}, \nu_{2},\nu_{3})^T_R$,  $ X_R=( X_{1}, X_{2} ,X_{3})^T_R$, the  neutrino mass terms derived from \eqref{eq_SSterm} are  written in the following ISS form  \cite{ParticleDataGroup:2020ssz}: 
\begin{align}
	- \mathcal{L}^{\nu}_{\mathrm{mass}}
	&\equiv\frac{1}{2}\left(\overline{(\nu_L)^c},\; \overline{\nu_R}, \; \overline{X_R}\right)
	\mathcal{M}^{\nu} \begin{pmatrix}
		\nu_L	\\
		\left( \nu_R\right)^c	\\
		\left( X_R\right)^c
	\end{pmatrix}
	+ \mathrm{h.c.},
	\crn  
	\mathcal{M}^{\nu}&= 
	\left(
	\begin{array}{cc}
		\mathcal{O}_{3\times3} & M_D^T \\
		M_D & M_{N} \\
	\end{array}
	\right), \;  M_D= \begin{pmatrix}
		m_D	\\
		\mathcal{O}_{3\times3}
	\end{pmatrix}, 
	\;  M_N=\left(
	\begin{array}{cc}
		\mathcal{O}_{3\times3}& M_R \\
		M^T_R & \mu_X \\
	\end{array}
	\right), 
	\label{eq_L0ISSnumass}	
\end{align}
where $\mathcal{O}_{3\times3}$ is a zero matrix  and $m_D=Y^{\nu}\times v_2/\sqrt{2}$. The analytic form of the Dirac mass matrix was chosen generally following Ref. \cite{Casas:2001sr}.

The total unitary mixing matrix $U^\nu$  is defined  as follows 
\begin{align}
	U^{\nu T}\mathcal{M}^{\nu}U^{\nu}=\hat{\mathcal{M}}^{\nu}=\mathrm{diag}(m_{n_1},m_{n_2},m_{n_3}, m_{n_4},..., m_{n_{9}})\equiv \mathrm{diag} \left( \hat{m}_{\nu},\; \hat{M}_N\right), \label{eq_dUnu}	
\end{align}
where   $m_{n_i}$ ($i=1,2,...,9$) are  eigenvalues of the $9$ mass eigenstates $n_{iL}$,  including three  light active neutrinos $n_{aL}$ ($a=1,2,3$) with mass matrix $\hat{m}_{\nu}$ and six other heavy neutrinos with mass matrix $\hat{M}_N$. 
The relation between the flavor  and mass eigenstates are
\begin{equation}
	\begin{pmatrix}
		\nu_L	\\
		\left( \nu_R\right)^c\\
		\left( X_R\right)^c	
	\end{pmatrix}=U^{\nu} n_L, \hs \mathrm{and} \; \begin{pmatrix}
		\left( \nu_L\right)^c	\\
		\nu_R	\\
		X_R	
	\end{pmatrix}=U^{\nu*}  n_R, \label{eq_Nutrans}
\end{equation}
where $n_L\equiv(n_{1L},n_{2L},...,n_{9L})^T$ and $n_R=\left(n_L\right)^c$. 
%
The neutrino mixing matrix is parameterized in the following form:
\begin{equation} 
	U^{\nu}= \left(
	\begin{array}{cc}
		\left(	I_3-\frac{1}{2}RR^{\dagger} \right) U_{\mathrm{PMNS}} & RV \\
		-R^\dagger U_{\mathrm{PMNS}} & \left(I_6 -\frac{1}{2}R^{\dagger} R\right)V \\
	\end{array}
	\right)  +\mathcal{O}(R^3), 
	\label{eq_Unu0}	
\end{equation}
where   $U_{\mathrm{PMNS}}$  is the  $3\times3$  Pontecorvo-Maki-Nakagawa-Sakata (PMNS) matrix \cite{Pontecorvo:1957cp, Maki:1962mu},  $V$ is a $6\times 6$ unitary matrix,  and  $R$ is a $3\times 6$ matrix satisfying $|R_{aI}|\ll1$ for all $a=1,2,3$ and $I=1,2,...,6$. 
In the ISS framework we considered  here, $m_D$ is parameterized in terms of many free parameters, hence it is convenient to choose a simple form of  $\mu_X=\mu_0 I_3$ and    $M_R=\hat{M}_R= M_0I_3$ \cite{Arganda:2014dta, Thao:2017qtn}. The formulas of  $m_D$  and mixing parameters  are  \cite{Casas:2001sr} 
\begin{equation}
	m_D= M_0\sqrt{\hat{x}_\nu}  U^{\dagger}_{\mathrm{PMNS}},\;  R  \simeq \left(0,\;  U_{\mathrm{PMNS}}\hat{x}_\nu^{1/2} \right), \; \hat{x}_\nu\equiv \frac{\hat{m}_\nu}{\mu_0}, \label{eq_mDiss}
\end{equation}
where  max$[\left|\left(\hat{x}_{\nu}\right)_{aa} \right|]\ll1$ for all $a=1,2,3$.  

The ISS conditions $|\hat{m}_{\nu}|\ll |\mu_0|\ll |m_D|\ll M_0$ so that $\frac{\sqrt{\mu_0 \hat{m}_{\nu}}}{M_0}\simeq0$,  the mixing matrix and Majorana mass term are 
\begin{align}
	\hat{M}_N= \left(\begin{matrix}
		\hat{M}_R	& 0 \\ 
		0	& \hat{M}_R
	\end{matrix} \right)\simeq  M_0I_6 , 
	\;  V\simeq \dfrac{1}{\sqrt{2}}
	\left(\begin{matrix}
		-iI_3 	& I_3   \\ 
		iI_3 	& I_3 
	\end{matrix} \right), \label{eq_UNiss}	
\end{align}
which give $V^*\hat{M}_NV^{\dagger} \simeq M_{N}$, i.e.,  $m_{n_i}\simeq M_0$ for all $i=4,5,...,9$. 

\subsection{Analytic formulas for one-loop contributions to $a_{e_a}$ and Br$(e_b\to e_a\gamma)$}

We note that except the contributions from  ISS neutrino couplings with singly charged Higgs bosons given in Eq. \eqref{eq_scHigg}, all other contributions are the same results as  those discussed in the case of Dirac active neutrinos mentioned above.  Hence, we just focus here on the Higgs contributions relating to the couplings with ISS neutrinos. The relevant couplings  are listed in  the following Lagrangian
\begin{align}
	\label{eq_g2Lagrangian}
	\mathcal{L} =& -\frac{g}{\sqrt{2} m_W}\sum_{k=1}^2\sum_{a=1}^3 \sum_{i=1}^{9} \overline{n_i} \left[ 	\lambda^{L,k}_{ia}  P_L +\lambda^{R,k}_{ia}P_R \right] e_a h^+_k   
	+\sum_{a=1}^3\sum_{i=1}^{9}\frac{g}{\sqrt{2}} U^{\nu*}_{ai} \overline{n_{i}}\gamma^{\mu}P_L e_aW^+_{\mu}
	\crn& +\mathrm{h.c.},
\end{align}
where
\begin{align}
	\label{eq_lakLR}
	\lambda^{L,1}_{ia}&= \sum_{I=1}^6 t_\beta^{-1}M_{D,Ia}c_{\alpha}U^{\nu}_{(I+3)i} \simeq \frac{t_\beta^{-1} c_{\alpha}}{\sqrt{2}}\times \left[\begin{array}{cc}
		0,	& \quad i\leq 3 \\
		-iM_0\left( U^*_{\mathrm{PMNS}}\hat{x}^{1/2}_{\nu} \right)_{a(i-3)}	& \quad  3<i\leq6  \\
		M_0\left(U^*_{\mathrm{PMNS}}\hat{x}^{1/2}_{\nu}\right)_{a(i-6)}	& \quad i\geq 7 
	\end{array}\right., 
	\crn \lambda^{L,2}_{ia}& \simeq	\lambda^{L,1}_{ia}t_{\alpha},
	\crn	\lambda^{R,1}_{ia}&= m_{e_a}t_{\beta} c_{\alpha} U^{\nu*}_{ai} - \sum_{I=1}^6 \frac{v}{\sqrt{2}}  s_{\alpha} Y^{\sigma}_{Ia}U^{\nu*}_{(I+3)i} 
	\crn &\simeq \left[\begin{array}{cc}
		m_{e_a}t_{\beta} c_{\alpha} \left[ U^*_{\mathrm{PMNS}}\left( I_3 -\frac{1}{2}\hat{x}_{\nu} \right) \right]_{ai} + \frac{vs_{\alpha}}{\sqrt{2}} \left( Y^{\sigma T} \hat{x}_{\nu}^{1/2}\right)_{ai}	& \quad i\leq 3 \\
		\frac{-i}{\sqrt{2}}m_{e_a}t_{\beta} c_{\alpha} \left( U^*_{\mathrm{PMNS}}\hat{x}_{\nu}^{1/2} \right)_{a(i-3)}+	\frac{ivs_{\alpha}}{2}\left[Y^{\sigma T} \left(I_3- \frac{1}{2}\hat{x}_{\nu} \right)\right]_{a(i-3)}	& \quad  4 \leq i<7\\
		\frac{1}{\sqrt{2}}m_{e_a}t_{\beta} c_{\alpha} \left( U^*_{\mathrm{PMNS}}\hat{x}_{\nu}^{1/2} \right)_{a(i-6)} 	-\frac{vs_{\alpha}}{2}\left[Y^{\sigma T} \left(I_3- \frac{1}{2}\hat{x}_{\nu} \right)\right]_{a(i-6)}		& \quad i \geq 7 
	\end{array}\right.,
	\crn	\lambda^{R,2}_{ia}&=	\lambda^{R,1}_{ia}\left[ s_{\alpha}\to - c_{\alpha}, \; c_{\alpha} \to s_{\alpha}\right].
\end{align}	

The branching ratios of the cLFV decays are formulated as follows \cite{Lavoura:2003xp, Hue:2017lak, Crivellin:2018qmi}:
\begin{align}
	\label{eq_brebaga}
	\mathrm{Br}(e_b\to e_a\gamma)= \frac{48\pi^2}{G_F^2 m_b^2}\left( \left| c_{(ab)R}\right|^2 + \left| c_{(ba)R}\right|^2\right) \mathrm{Br}(e_b\to e_a \overline{\nu_a}\nu_b),
\end{align}
where $G_F=g^2/(4\sqrt{2}m_W^2)$, Br$(\mu\to e \overline{\nu_e}\nu_{\mu})\simeq 1$,  Br$(\tau\to e \overline{\nu_e} \nu_{\tau}) \simeq 0.1782$, Br$(\tau\to \mu \overline{\nu_\mu}\nu_{\tau})\simeq 0.1739$ \cite{ParticleDataGroup:2020ssz},  and 
\begin{align}
	\label{eq_cabR}	
	c_{(ab)R}&= \sum_{k=1}^2 c_{(ab)R} \left(h^\pm\right) + c_{(ab)R}(W), \quad   c_{(ba)R}= c_{(ab)R}[a \to b,\; b\to a], 
	\crn c_{(ab)R}\left(h^\pm\right) &= \sum_{k=1}^2 c_{(ab)R} \left(h^\pm_k\right),
	\crn 	c_{(ab)R} \left(h^\pm_k\right)& = \frac{g^2e\;}{32 \pi^2 m^2_Wm^2_{h^\pm_k} }  
	\crn&\times\sum_{i=1}^{9}  \left[ \lambda^{L,k*}_{ia } \lambda^{R,k}_{ib }m_{n_i} f_{\Phi}(x_{i,k}) + \left( m_{e_b} \lambda^{L,k*}_{ia } \lambda^{L,k}_{ib } + m_{e_a} \lambda^{R,k*}_{ia } \lambda^{R,k}_{ib }\right)  \tilde{f}_{\Phi}(x_{i,k}) \right],
\end{align}
with $x_{i,k}\equiv m^2_{n_i}/m^2_{h^\pm_k}$. 

Up to the order $\mathcal{O}(R^2)$ of the neutrino mixing matrix given in Eq. \eqref{eq_Unu0}, the non-zero one-loop contributions relating to $h^\pm_{1,2}$ are
\begin{align}
	\label{eq_cabRX}	
	c_{(ab)R} \left(h^\pm_1\right) &=\frac{e G_F m_{e_b}}{4\sqrt{2} \pi^2} 
	\crn&\times \left\{  \left[ c^2_{\alpha} \left(U_{\mathrm{PMNS}}\hat{x}_{\nu}U_{\mathrm{PMNS}}^\dagger \right)_{ab}  - \frac{vt_{\beta}^{-1}c_{\alpha}s_{\alpha}}{m_{e_b} \sqrt{2}} \left(U_{\mathrm{PMNS}}\hat{x}_{\nu}^{1/2}Y^{\sigma}\right)_{ab}   \right]x_1f_{\Phi}(x_1)
	\right.\crn&  \quad +  t_{\beta}^{-2} c^2_{\alpha}\left(U_{\mathrm{PMNS}}\hat{x}_{\nu}U_{\mathrm{PMNS}}^\dagger\right)_{ab}   x_1 \tilde{f}_{\Phi}(x_1)
	\crn & \quad+ \frac{m^2_{e_a} t_{\beta}^2c_{\alpha}^2}{m^2_{h^\pm_1}} \left[ \frac{\delta_{ab}}{24} -  \left(U_{\mathrm{PMNS}}\hat{x}_{\nu}U_{\mathrm{PMNS}}^\dagger\right)_{ab}    \left( \frac{1}{24} -\tilde{f}_{\Phi}(x_1) \right) \right]
	\crn & \quad+ \frac{m_{e_a}v^2s_{\alpha}^2}{2 m_{e_b} m^2_{h^\pm_1}} \left[\left(Y^{\sigma \dagger}\hat{x}_{\nu}Y^{\sigma }\right)_{ab}   \left(  \frac{1}{24} -\tilde{f}_{\Phi}(x_1)  \right)  + \left(Y^{\sigma\dagger}Y^{\sigma}\right)_{ab} \tilde{f}_{\Phi}(x_1)\right]
	\crn& \quad \left. +  \frac{vm_{e_a}t_{\beta} s_{2\alpha}}{2 \sqrt{2} m^2_{h^\pm_1}} \left( \frac{1}{24}- \tilde{f}_{\Phi}(x_1)  \right) \left[ \frac{m_{e_a}}{m_{e_b}} \left(U_{\mathrm{PMNS}}\hat{x}_{\nu}^{1/2}Y^{\sigma} \right)_{ab} +  \left(U_{\mathrm{PMNS}}\hat{x}_{\nu}^{1/2}Y^{\sigma}\right)^*_{ba}\right] 
	\right\},
	\crn c_{(ab)R}(h^\pm_2)&=c_{(ab)R}(h^\pm_1) \left[ x_1 \to\; x_2,\;s_{\alpha}\to - c_{\alpha}, \; c_{\alpha} \to s_{\alpha}\right],
	\crn 	c_{(ab)R}(W)& \simeq\frac{e G_Fm_{e_b}}{4\sqrt{2} \pi^2}   \left[  -\frac{5\delta_{ab}}{12} +  \left(U_{\mathrm{PMNS}}\hat{x}_{\nu}U_{\mathrm{PMNS}}^\dagger \right)_{ab} \times \left(\tilde{f}_V \left(x_W\right)+ \frac{5}{12}\right) \right].
\end{align}
The one-loop contributions from the $h^\pm_{1,2}$ exchanges to $a_{e_a}$ are: 
\begin{align}
	\label{eq_Hpm1}
	a_{e_a}(h^\pm_1)&=-\frac{4m_{e_a}}{e} \mathrm{Re}[c_{(aa)R}(h^\pm_1)], 
	\crn a_{e_a}(h^\pm_2)&= a_{\mu}(h^\pm_1) \left[ x_1 \to\; x_2,\;s_{\alpha}\to - c_{\alpha}, \; c_{\alpha} \to s_{\alpha}\right],
	\crn a_{e_a}(h^\pm)&=a_{e_a}(h^\pm_1) +a_{e_a}(h^\pm_2),
\end{align}
where $x_k=M_0^2/m^2_{h^\pm_k}$ and $a_{e_a}(h^\pm)$ is the total contribution from these two Higgs bosons. We note that all Yukawa couplings considered in this work are assumed to be real.  As a result, Eq. \eqref{eq_cabRX} shows that only the Dirac phase results in the tiny values of  $\mathrm{Im}[c_{aa}]$,  implying very suppressed values of electric dipole moments $d_{e_a}\equiv -2\; \mathrm{Im}[c_{aa}]$ \cite{Crivellin:2018qmi}, consistent with experimental constraints \cite{Muong-2:2008ebm, Roussy:2022cmp}. This conclusion is also confirmed by our numerical investigation, namely $|d_{e,\mu}|<10^{-44}[e\; \mathrm{cm}]$, therefore we will not discuss  it further.

For qualitative estimation, the main contribution to $a_{e_a}(h^\pm)$ is the  chirally-enhanced part \cite{Crivellin:2018qmi}. From Eq. \eqref{eq_cabRX}, this part is determined  as follows
\begin{equation}\label{eq_aea0}
	a_{e_a,0}=	\frac{G_Fm^2_{e_a}}{\sqrt{2}\pi^2}
	\times \mathrm{Re}\left\{  \left[ \frac{vt_{\beta}^{-1}c_{\alpha}s_{\alpha}}{\sqrt{2}m_{e_a}}U_{\mathrm{PMNS}} \hat{x}_{\nu}^{1/2}  Y^{\sigma}\right]_{aa} \left[ x_1f_{\Phi}(x_1) - x_2f_{\Phi}(x_2)\right]
	\right\}.
\end{equation}
But this term may also give large contributions to $c_{(ab)R}$ and $c_{(ba)R}$, which may  be excluded by the cLFV constraints  with $b\neq a$. To avoid this problem, we start  from the diagonal form of  $c_{(ab)R,0}$  as follows 
\begin{equation}\label{eq_cab0}
	c_{(ab)R,0}\varpropto\;   \left[ U_{\mathrm{PMNS}} \hat{x}_{\nu}^{1/2}  Y^{\sigma} \right]_{ab} \varpropto\;  \delta_{ab}.
\end{equation}
Correspondingly,   the formula of  	$a_{e_a,0}$ is proportional to a diagonal matrix $Y^d$ satisfying: 
\begin{align} \label{eq_defYd}
	&U_{\mathrm{PMNS}} \times \mathrm{diag}\left( \frac{m_{n_1}}{m_{n_3}},\; \frac{m_{n_2}}{m_{n_3}},\;1\right)^{1/2}  Y^{\sigma}=	 Y^d  \equiv \mathrm{diag} \left(Y^d_{11}, \; Y^d_{22}, Y^d_{33}\right), 
\end{align}
Hence, the diagonal entries will give main contributions to $a_{e_a}$, namely  
\begin{equation}\label{eq_aea01}
	a_{e_a,0}=	\frac{G_Fm^2_{e_a} \sqrt{x_0}}{\sqrt{2}\pi^2}
	\times \mathrm{Re}\left[ \frac{vt_{\beta}^{-1}c_{\alpha}s_{\alpha}}{\sqrt{2}m_{e_a}}Y^d\right]_{aa} \left[ x_1f_{\Phi}(x_1) - x_2f_{\Phi}(x_2)\right].
\end{equation}
Then $	a_{e_a,0}$ may be large, provided that  $t_{\beta}$ should not too large,  $t_1\neq t_2$, and the following quantities are large enough:  $x_0$, $|s_{2\alpha}=2s_{\alpha} c_{\alpha}|$, and  $|Y^d_{aa}|$ with $a=1,2,3$. In contrast, cLFV amplitudes do not get any contributions from $c_{(ab) R,0}$. Numerical investigations will be done to check this conclusion.  For simplicity, we use the approximation that all tiny contributions are ignored in the numerical results.  Namely,  $\Delta a_{e_a}\equiv a_{e_a}(h^\pm)$ given in Eq. \eqref{eq_cabRX}, instead of the total formula  given in Eq. \eqref{eq_defDamu341}. In contrast, the one-loop contribution from  $W$  must be included in the formulas of Br$(e_b\to e_a \gamma)$.   This conclusion was confirmed based on the qualitative estimation discussed above. The numerical checks have been performed, which are  consistent with previous discussions on 3-3-1 models \cite{Hue:2021xap, Hong:2022xjg}.  

Although our numerical investigation will consider only the experimental constraint given in Eq. \eqref{eq_dae} allowing only positive values of $\Delta a_{e}^{\mathrm{NP}}$  around 1$\sigma$ constraint, the experimental result with negative $\Delta a_{e}^{\mathrm{NP}}$ in Ref. \cite{Parker:2018vye} can be explained similarly because the sign of the main contribution in Eq. \eqref{eq_aea01} depend precisely on the signs of $Y^d s_{\alpha}c_{\alpha}$.    

\section{ \label{sec_numerical} Numerical discussion}

We will use the best-fit values  of the neutrino osculation data \cite{ParticleDataGroup:2020ssz} corresponding to  the normal order (NO) scheme  with $m_{n_1}<m_{n_2}<m_{n_3}$, namely 
\begin{align}
	\label{eq_d2mijNO}
	&s^2_{12}=0.32,\;   s^2_{23}= 0.547,\; s^2_{13}= 0.0216 ,\;  \delta= 218 \;[\mathrm{Deg}] , 
	\crn &\Delta m^2_{21}=7.55\times 10^{-5} [\mathrm{eV}^2], \;
	\Delta m^2_{32}=2.424\times 10^{-3} [\mathrm{eV}^2].
\end{align}
The active mixing matrix and neutrino masses are determined  as follows 
\begin{align}
	\label{eq_NO1}
	\hat{m}_{\nu}&= \left( \hat{m}^2_{\nu}\right)^{1/2}= \mathrm{diag} \left( m_{n_1}, \; \sqrt{m^2_{n_1} +\Delta m^2_{21}},\; \sqrt{m^2_{n_1} +\Delta m^2_{21} +\Delta m^2_{32}} \right),
	\crn U_{\mathrm{PMNS}}&=\left(
	\begin{array}{ccc}
		c_{12} c_{13} & c_{13} s_{12} & s_{13} e^{-i \delta } \\
		-c_{23} s_{12}-c_{12} s_{13} s_{23} e^{i \delta } & c_{12}c_{23}-s_{12} s_{13} s_{23} e^{i \delta } & c_{13} s_{23} \\
		s_{12} s_{23}-c_{12} c_{23} s_{13} e^{i \delta } & -c_{23} s_{12} e^{i \delta } s_{13}-c_{12} s_{23} & c_{13} c_{23} \\
	\end{array}
	\right).
\end{align}
This choice of active neutrino masses also satisfies the constraint  from Plank2018 \cite{Planck:2018vyg}, $	\sum_{i=a}^{3}m_{n_a}\leq 0.12\; \mathrm{eV}$.

The non-unitary of the active neutrino mixing matrix $\left(	I_3-\frac{1}{2}RR^{\dagger} \right) U_{\mathrm{PMNS}}$ is constrained by other phenomenology such as electroweak precision~\cite{Fernandez-Martinez:2016lgt,  Agostinho:2017wfs, Coutinho:2019aiy}, leading to a very strict constraint of $\eta\equiv	\frac{1}{2}\left| RR^{\dagger}\right| \varpropto\;   \hat{x}_\nu$ in the ISS framework \cite{Mondal:2021vou, Biggio:2019eeo, Escribano:2021css}.  The reasonable constraint of the non-unitary part of the neutrino mixing matrix is  as follows 
\begin{equation}\label{eq_x0}
	x_0\equiv \frac{m_{n_3}}{\mu_0}\leq 10^{-3}.
\end{equation}
The other well-known numerical parameters are \cite{ParticleDataGroup:2020ssz} 
\begin{align}
	\label{eq_ex}
	g &=0.652,\; G_F=1.1664\times 10^{-5}\; \mathrm{GeV},\; s^2_{W}=0.231,\;m_W=80.385 \; \mathrm{GeV}, \crn
	m_e&=5\times 10^{-4} \;\mathrm{GeV},\; m_{\mu}=0.105 \;\mathrm{GeV}, \; m_{\tau}=1.776  \; \mathrm{GeV}. 
\end{align}
For the free parameters of the 341ISS model, the numerical scanning ranges are 
\begin{align} \label{eq_scanP0}
	M_0&\in \left[ 0.1,\;10\right]\; \mathrm{TeV},\;  m_{h^\pm_{1,2}}\in \left[0.8,\;10\right]\; \mathrm{TeV},
	\crn t_{\beta} &\in \left[ 0.3,50\right],\; x_0 \in \left[ 10^{-6},10^{-3}\right],\;s_{\alpha} \in  \left[-1,1\right], \;|Y^d_{ab}|\leq 3.5\; \forall a,b=1,2,3.  
\end{align}
In addition, we will fix $m_{n_1}=0.01$ eV, and check the perturbative limit of all  Yukawa couplings $Y^{\sigma}$ and $Y^{\nu}$, namely $|Y^\sigma_{ab}|,|Y^\nu_{ab}|\le 3.5$ must be satisfied.  As we discussed above, the diagonal form of $Y^d$ will allow large $(g-2)_{e_a}$ anomalies and small Br$(e_b \to e_a \gamma)$ satisfying the recent experimental constraints. In the numerical investigation, we will consider the general case that $Y^d_{ab}\neq 0$. The allowed regions of parameters we imply below must guarantee both the experimental data of $1\sigma$ ranges of  $(g-2)_{e,\mu}$, the cLFV constraints of Br$(e_b\to e_a \gamma)$ , and the  constraint of  the decay $Z\to \mu^+\mu^-$ using the analytic formulas and experimental data given in appendix \ref{app_Zmm}. 

Now, we  consider some particular different choices of zero entries of  $Y^d$. The allowed regions of the parameter space predict some interesting  properties. First, even with only $Y^d_{11},Y^d_{22}\neq0$, the allowed regions satisfying two $(g-2)_{e,\mu}$ data are still strictly constrained by  the cLFV decay rate Br$(\mu \to e\gamma)<4.2 \times 10^{-13}$, namely  the constraints  $|Y^d_{12}|,|Y^d_{21}|<10^{-4}$ must be guaranteed. Therefore, we will fix $|Y^d_{12}|= |Y^d_{21}|=0$ as the default values in our numerical investigations.  Second, we give comments on the three following cases:
\begin{enumerate}
	\item $Y^d_{23}=Y^d_{32} =Y^d_{13}=Y^d_{31}=0$, which is the case of $Y^d$ being diagonal, as given in Eq. \eqref{eq_defYd}.  The allowed regions  of non-zero entries of $Y^d$ are $0.02\leq |Y^d_{11}|\le 0.13$, $0.9\le |Y^d_{22}|\le2.5$, and $|Y^d_{33}|\leq 2.67$,  see the left panel of Fig. \ref{fig_fbeta123}. 
	\begin{figure}[ht]
		\centering\begin{tabular}{cc}
			\includegraphics[width=7.5cm]{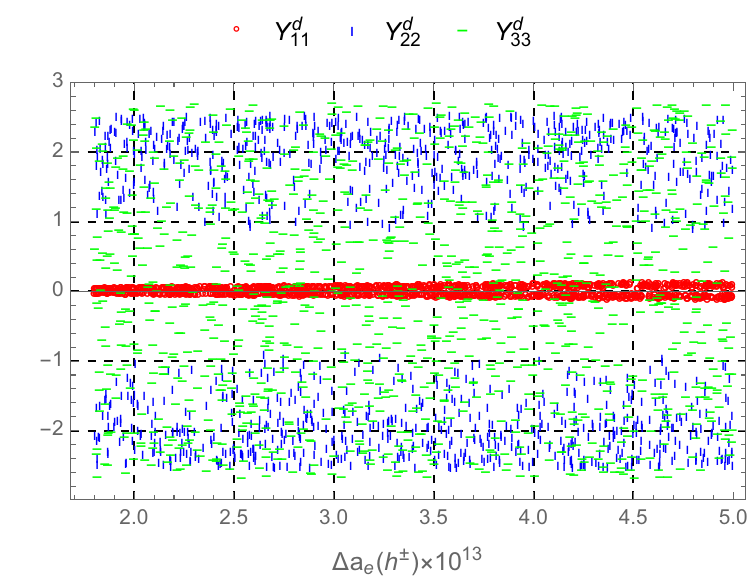}&
			\includegraphics[width=7.5cm]{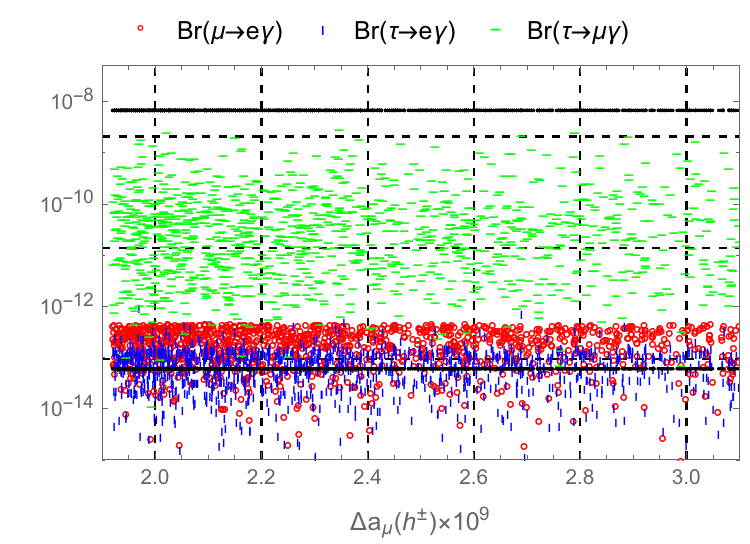}\\ 
		\end{tabular}
		\caption{ The  correlations between $\Delta a_{\mu}(h^\pm)$   vs $Y^d_{aa}$ (left panel) and $\Delta a_{e}(h^\pm)$   vs Br$(e_b \to e_a \gamma)$ (right panel) in the case  $Y^d_{23}=Y^d_{32} =Y^d_{13}=Y^d_{31}=0$.  The two black line show the values of $6.9\times 10^{-9}$ and $6\times 10^{-14}$ corresponding to the future experimental sensitivities of cLFV decays $e_b\to e_a\gamma$ mentioned in the introduction. }\label{fig_fbeta123}
	\end{figure}
	In this allowed region, Br$(\tau\to \mu \gamma)$ can reach the order of $\mathcal{O}(10^{-9})$, but predicts suppressed branching ratios  Br$(\tau\to e\gamma)<10^{-12}$, see the right panel of Fig. \ref{fig_fbeta123}.    We can see that Br$(\mu \to e \gamma)$ can reach the recent constraint of $\mathcal{O}(10^{-13})$ even when $Y^d_{12}=Y^d_{21}=0$.  This property is different from two other cLFV decays of $\tau$, which  are  suppressed if  $Y^d_{33}=0$ is fixed, namely  Br$(\tau\to \mu \gamma)<10^{-11}$.    In addition, there exist regions that allow both  small values of Br$(\mu \to e \gamma) \sim \mathcal{O}(10^{-14})$ corresponding to the future experimental sensitivity and  large $\Delta a_{e,\mu}$. In conclusion, we confirm that the formula of $a_{e,0}$ given in Eq. \eqref{eq_aea0} is the main one-loop contribution to $(g-2)_{e_a}$ data and cLFV decay amplitudes. Hence,  the diagonal form of $Y^d$ results in  small cLFV decay rates, even though they get contributions from other terms  in Eq. \eqref{eq_cabRX}.     
	
	\item $Y^d_{32}=Y^d_{23}=Y^d_{33}=0$ while  $Y^d_{31},Y^d_{31} \neq 0$. The two cLFV rates Br$(\tau \to e\gamma)$ and Br$(\mu \to e\gamma)$ can reach recent experimental upper bound, while  Br$(\tau\to \mu \gamma)< 3\times 10^{-12}$.   The allowed ranges of  $Y^d_{11}$ and $Y^d_{22}$ are almost unchanged. The constraints of $Y^d_{13,31}$  are $|Y^d_{13}|<0.67$ and $|Y^d_{31}|<0.22$.  Illustrations for  relations between entries of $Y^d$  and Br$(\tau\to e\gamma)$ are presented in  Fig. \ref{fig_fbeta13}. 
	\begin{figure}[ht]
		\centering\begin{tabular}{cc}
			\includegraphics[width=7.5cm]{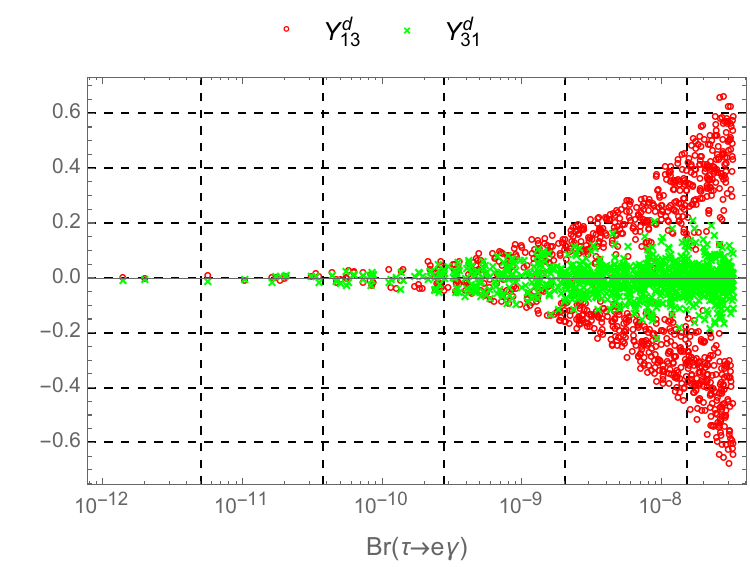}&
			\includegraphics[width=7.5cm]{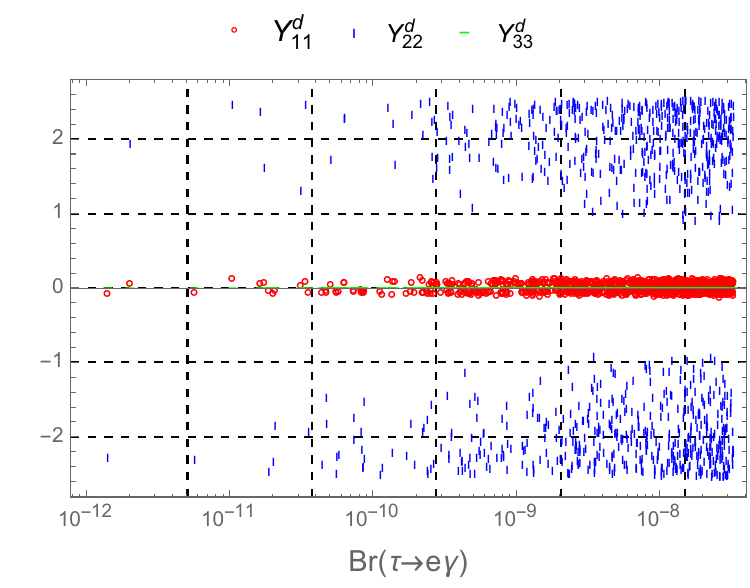}\\ 
		\end{tabular}
		\caption{ The  correlations between Br$(\tau \to e\gamma)$   vs $Y^d_{13,31}$ (left panel) and    $Y^d_{aa}$  (right panel) in the second  case  $Y^d_{23}=Y^d_{32} =Y^d_{33}=0$.  }\label{fig_fbeta13}
	\end{figure}
	We conclude that  Br$(\tau \to e \gamma)$ depends  strongly on $Y^d_{13}$ and $Y^d_{31}$. In contrast, the allowed ranges of $Y^d_{11}$ and $Y^d_{22}$ are mainly arising from the $(g-2)$ data.

	\item $Y^d_{13} =Y^d_{31}= Y^d_{33} =0$ while $Y^d_{32}, Y^d_{2,3} \neq0$. Two cLFV decays Br$(\mu\to e \gamma)$ and Br$(\tau \to \mu\gamma)$ can reach recent experimental  bounds, but very suppressed Br$(\tau\to e\gamma)<10^{-12}$. The constraints of the non-zero entries of $Y^d$  are $|Y^d_{23,32}|\leq 1$. Illustrations between Br$(\tau \to \mu \gamma)$ vs. $Y^d_{23,32}$ and $a_{\mu,e}(h^\pm)$ are similar to the case 2, hence we do not show explicitly here. 
\end{enumerate}

We comment  here on some properties of entries of $Y^d$ derived  from studying three particular cases mentioned above.   Firstly, the diagonal  entries of $Y^d$  give main contributions to $(g-2)_{e_a}$ anomalies, while the non-zero entries $Y^d_{13,31}$ and  $Y^d_{23,32}$ affect strongly in the decay rates of Br$(\tau \to e\gamma)$ and Br$(\tau \to \mu\gamma)$, respectively. The Br$(\mu \to e\gamma)$ depends strongly on $Y^d_{12,21}$ and the combination of the remaining contributions. The recent experimental bound of Br$(\mu \to e\gamma)<4.2\times 10^{-13} $ results in tiny values $|Y^d_{12,21}|<10^{-4}$, hence we always fix these two entries being zeros.   We also emphasize that the negative sign of the experimental value of $\Delta a^{\mathrm{NP}}_e$ in Ref. \cite{Parker:2018vye}  can be explained by the sign of $Y^d_{22}$.

\begin{figure}[ht]
	\centering\begin{tabular}{cc}
		\includegraphics[width=7.5cm]{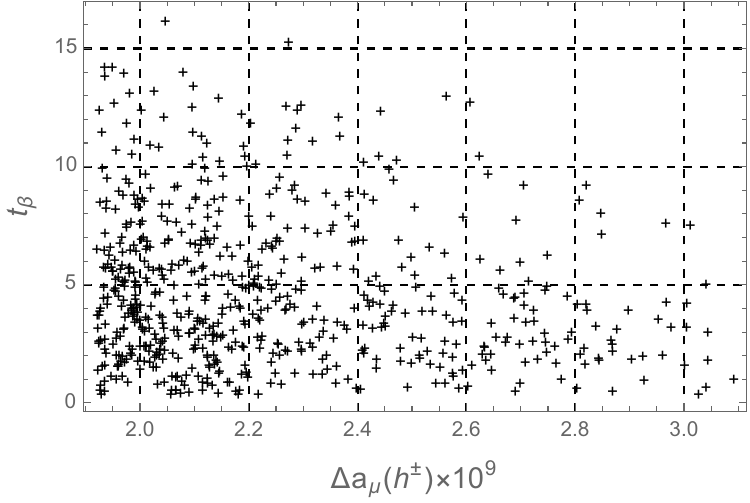}&
		\includegraphics[width=7.5cm]{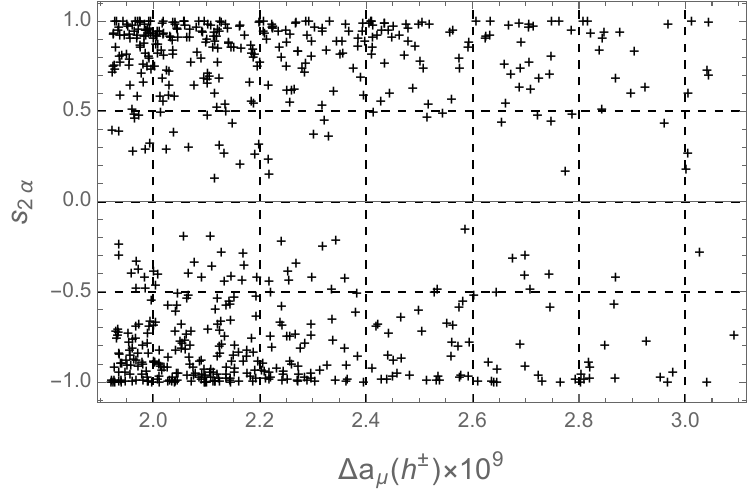}\\ 
		\includegraphics[width=7.5cm]{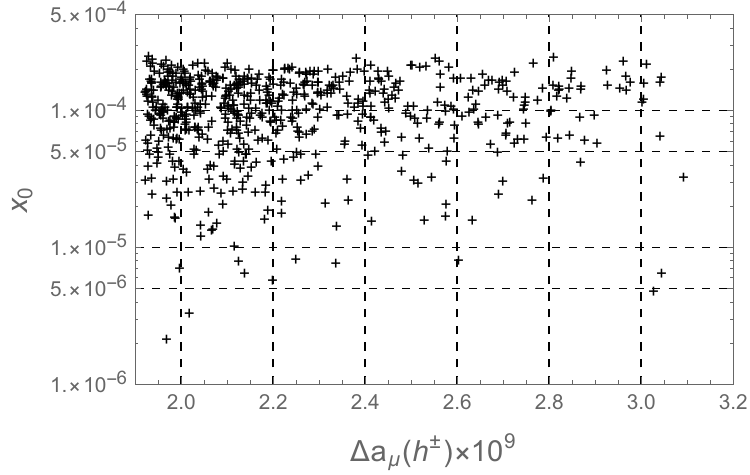}&
		\includegraphics[width=7.cm]{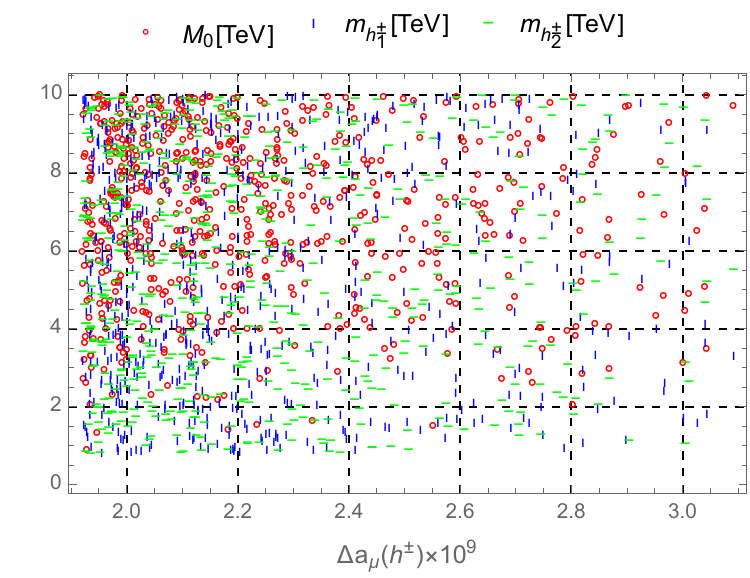}\\ 
		\includegraphics[width=7.cm]{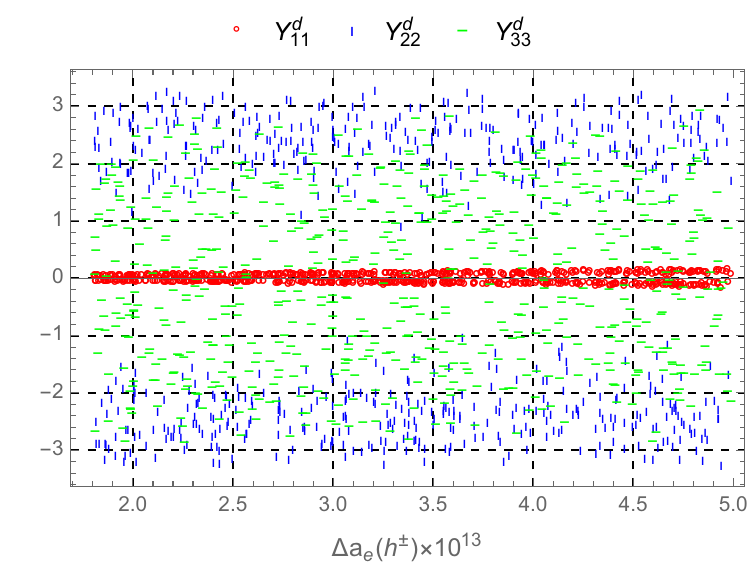}&
		\includegraphics[width=7.cm]{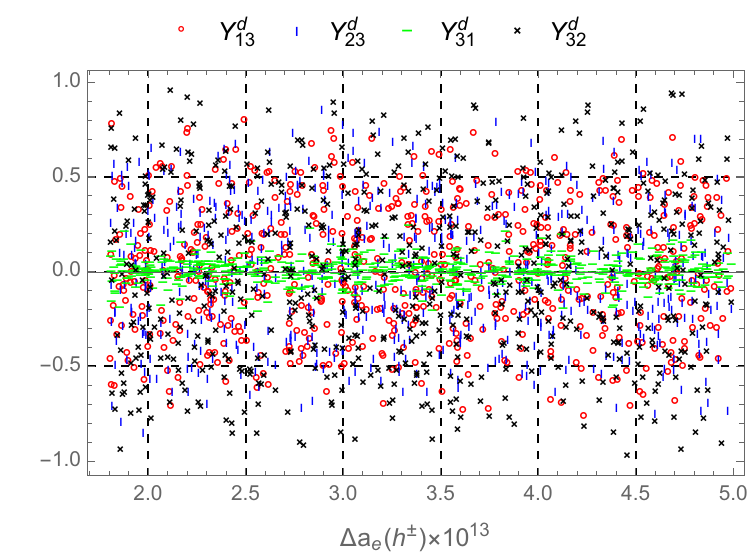}\\ 
	\end{tabular}
	\caption{ The  correlations between different free parameters  vs $\Delta a_{\mu}(h^\pm)$ with $Y_{21}=Y_{12}=0$.  }\label{fig_fbetaX2}
\end{figure}

In the final  illustration, we consider the more general case that the  only two zero entries of $Y^d$ are  $Y^d_{12} =Y^d_{21}=0$. The correlations of  important parameters vs. $\Delta a_{\mu} (h^\pm)$ are shown in Fig.~\ref{fig_fbetaX2}. 
The corresponding allowed ranges  of the free parameters are 
\begin{align}
	\label{eq_allowed} 
	t_{\beta } &\in \left[ 0.35,16.14\right],\; s_{\alpha }\in \left[-0.988,-0.08\right]\cup \left[ 0.065,0.996\right], \; x_0 \in \left[2.14\times 10^{-6},2.5 \times 10^{-4}\right], 
	\crn \; M_0 &\in \left[ 0.877,10\right]\; \text{[TeV]}, \; M_{1,2} \in \left[ 0.8,10\right]\; \text{[TeV]}, 
	\crn Y_{11}^d &\in \left[-0.17,-0.03\right]\cup \left[ 0.022, 0.164\right],\; Y_{22}^d\in \left[ -3.26,-1.03\right]\cup \left[ 0.909,3.275\right], \;   |Y_{33}^d| \leq 2.93, 
	\crn
	Y_{13}^d &\in\left[ -0.76,\; 0.8\right],\; 	Y_{31}^d \in\left[ -0.21,\; 0.25\right],\; Y_{23}^d \in \left[ -0.85,\; 0.86\right], \;	Y_{32}^d \in\left[ -0.97,\; 0.96\right].  
\end{align}
We can see that many allowed ranges are stricter than the scanning ones given in Eq. \eqref{eq_scanP0}.  For example, large $(g-2)_{e,\mu}$ requires small $t_\beta$ and large $x_0$ so that the new upper bound of $t_{\beta}$ and lower bound of $x_0$ are determined. As a result, in Fig. \ref{fig_fbetaX2}, the allowed regions of small $t_{\beta}$ and large $x_0$ are favored. Similarly, $a_{e_a,0}\varpropto s_{\alpha}c_{\alpha}=s_{2\alpha}/2 $, therefore, the allowed values of large $s_{2\alpha}$ close to 1 are supported. In the bottom left panel,  the allowed regions of $Y^d_{11,22}$ are the same as those predicted in Fig. \ref{fig_fbeta123}, in which all non-diagonal entries are zeros. This implies that  these regions are independent of   non-diagonal entries of $Y^d$. Instead, large values of these entries  favor the small $a_{\mu}(h^\pm)$, see the bottom right panel. As a result, small $(g-2)_{\mu}$ may predict large Br$(\tau\to \mu \gamma,\mu \gamma)$ and vice versa.  In addition to explain both $(g-2)_{e_a}$ data of $\Delta a_{\mu}(h^\pm)$,  the condition of $|m_{h^\pm_1}-m_{h^\pm_2}|\ge 523$ GeV is required, which is condition derived from $a_{e_a,0}$ that $x_1\neq x_2$. We emphasize that this model allows the existence of heavy charged Higgs bosons, which did not appear in some recent discussions on $(g-2)$ anomalies \cite{DelleRose:2020oaa, Botella:2020xzf}.

\begin{figure}[ht]
	\centering\begin{tabular}{cc}
		\includegraphics[width=7.cm]{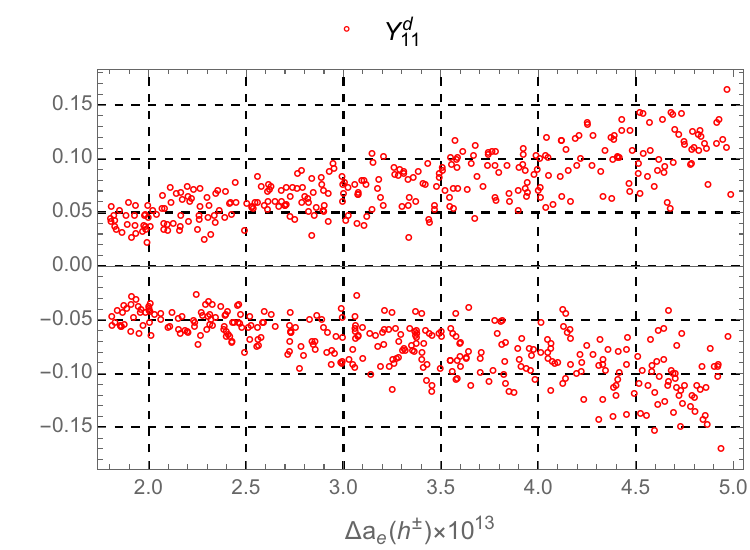}&
		\includegraphics[width=7.cm]{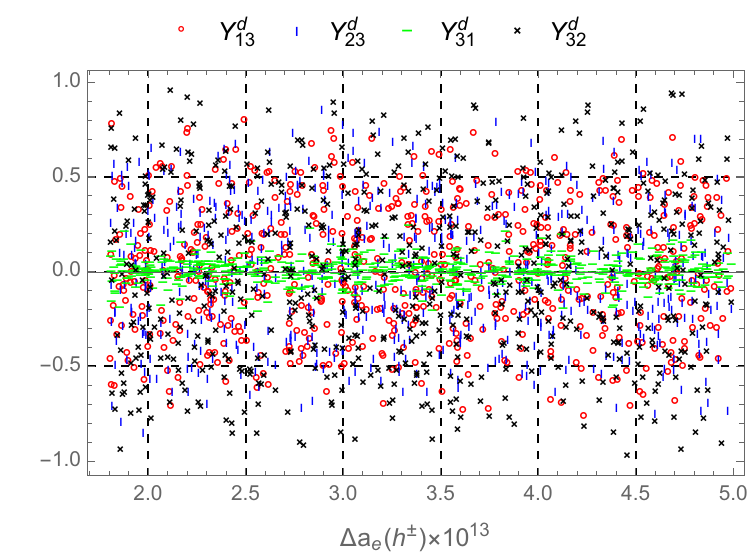}\\ 
	\end{tabular}
	\caption{ The  correlations between $Y^d_{ij}$  vs $\Delta a_{e}(h^\pm)$  with $Y_{21}=Y_{12}=0$.}\label{fig_aeX}
\end{figure}

The relation of  $Y^d_{ij}$  vs. $\Delta a_{e} (h^\pm)$  are shown in Fig.~\ref{fig_aeX}. 
In the left panel, $a_e(h^\pm)  \varpropto |Y^d_{11}|$  confirms the relation given in Eq. \eqref{eq_aea01}. The $a_{\mu}(h^\pm)  \varpropto\;  |Y^d_{22}|$  is not very clear, because many points are excluded by the perturbative limit. The right panel shows that $a_e(h^\pm)$ does not give any prediction on $Y^d_{ij}$ like the case of $a_{\mu}(h^\pm)$.

\begin{figure}[ht]
	\centering\begin{tabular}{cc}
		\includegraphics[width=7.cm]{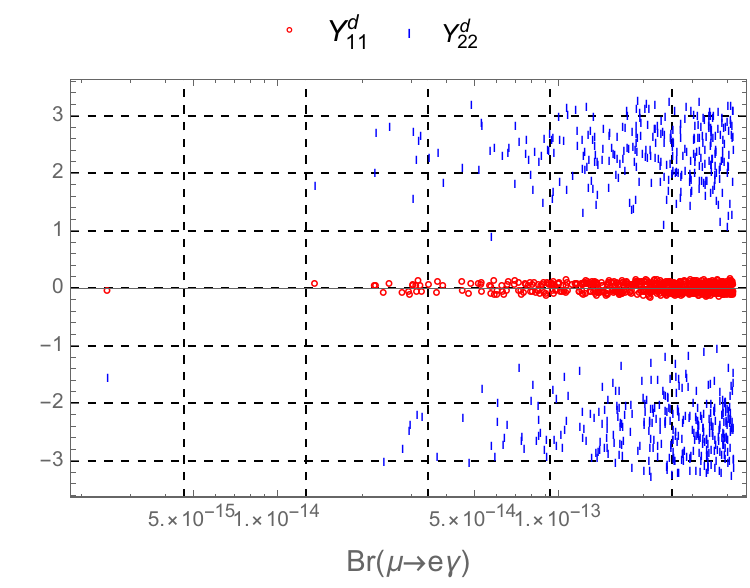}&
		\includegraphics[width=7.cm]{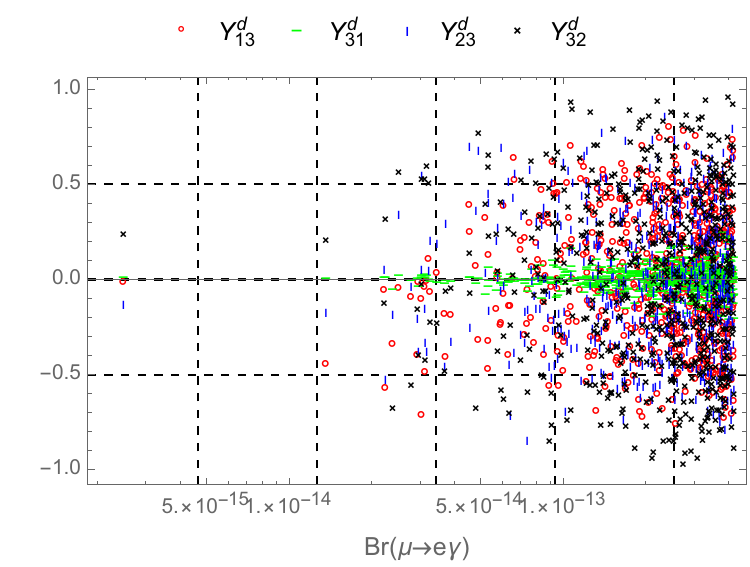}\\ 
	\end{tabular}
	\caption{ The  correlations between $Y^d_{ij}$ vs   Br$(\tau \to e\gamma,\mu\gamma)$ with $Y_{21}=Y_{12} =0$.  }\label{fig_Ydebaga}
\end{figure}

The correlations of  $Y^d_{ij}$ vs Br$(\mu\to e\gamma)$  are shown in Fig. \ref{fig_Ydebaga}.  
We see again that the constraints  of $Y^d_{ii}$ with $i=1,2$ are similar to those shown in the  left panel of Fig.~\ref{fig_fbeta123}.  This confirms the conclusion that the allowed ranges of $Y^d_{11,22}$ are mainly controlled by the $(g-2)_{e_a}$ data. In the right panel of Fig. \ref{fig_Ydebaga}, large Br$(\mu \to e \gamma)$ prefers large non-diagonal entries $Y^d_{ij}$ but the small values  are still allowed because of  the destructive correlations between contributions from different parameters. 

\begin{figure}[ht]
	\centering\begin{tabular}{cc}
		\includegraphics[width=7.cm]{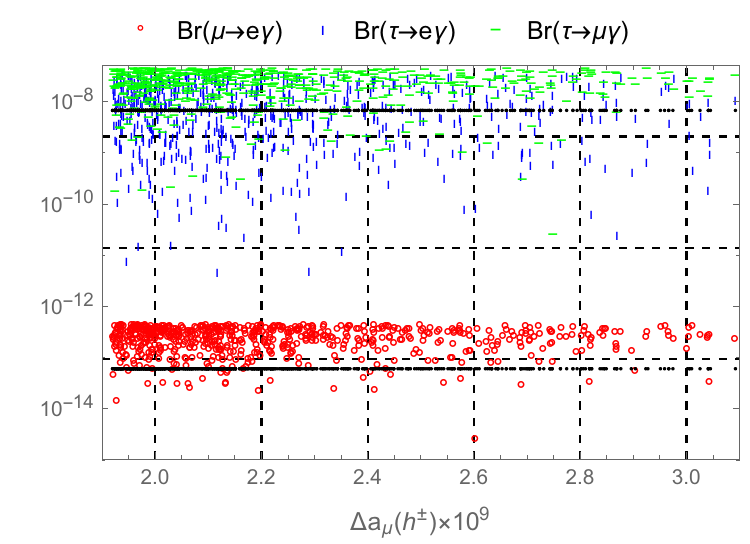}&
		\includegraphics[width=7.cm]{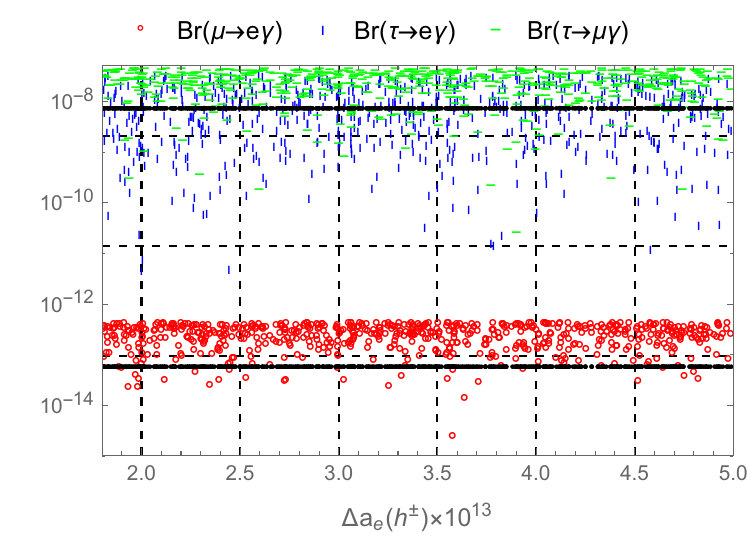}\\ 
	\end{tabular}
	\caption{ The  correlations between Br$(e_b \to e_a\gamma)$  vs $\Delta a_{e,\mu}(h^\pm)$ with $Y_{21}=Y_{12} =0$.  Two black lines in each panel present the future experimental sensitivities of Br$(e_b \to e_a \gamma) $ mentioned in the introduction. }\label{fig_aemcLFV}
\end{figure}

The correlations between $\Delta a_{e,\mu}$ and  Br$(e_b\to e_a\gamma)$ in the allowed regions of parameters listed in Eq. \eqref{eq_allowed}  are shown in Fig. \ref{fig_aemcLFV}. 
We can see that all allowed  values of $a_{\mu,e}(h^\pm)$ in the $1\sigma$ experimental ranges  still predict large Br$(e_b \to e_a \gamma)$ near the current upper experimental bounds. Therefore,  the future results of cLFV experiments, $(g-2)$ data, and neutrino oscillation will  give more strict constraints on the allowed regions of the parameter space. Note also that future experimental constraints on other cLFV decays such as $\mu\to 3e<10^{-16}$ \cite{Blondel:2013ia}  and $\mu-e$ conversion in nuclei \cite{Mu2e:2022ggl} will be more strict than those considered in this work, but in general  theoretical calculations on some other BSM in the presence of  "chiral enhancement"  show that they do not affect  strongly  on the allowed regions we discussed above, see for example  the left-right model \cite{Ashry:2022maw}. From the theoretical side, this can be explained by the reason that the relevant one-loop corrections originated from one-loop four-point diagrams are proportional to the products of four vertex factors, therefore their numerical values are more flexible  than the theoretical constraints from the cLFV decays $e_b\to e_a \gamma$. A more detailed investigation to predict the allowed regions corresponding to the future sensitivities  of all interesting cLFV decays will be done  in the future.

\section{\label{sec_con} Conclusion}
 We have discussed a solution to explain the recent experimental data of the $(g-2)_{e_a}$ anomalies in the 341ISS framework. We have constructed the Yukawa Lagrangian of leptons and Higgs potential obeying the  generalized lepton number $\mathcal{L}$  that keeps necessary terms generating  the ISS mechanism and large chirally-enhanced one-loop contributions to $(g-2)_{e_a}$ anomalies. Although the ISS mechanism may result in large cLFV decays $e_b\to e_a \gamma$, we have shown numerically that there always exist allowed regions of the parameter space guaranteeing these experimental bounds. In addition, these allowed regions will not be excluded totally if the future sensitivities of the cLFV experiments are updated, and no cLFV significations are found. The model can also  explain successfully the existence of  at least one of the cLFV decays $\tau \to \mu\gamma, e\gamma$,  or $\mu\to e\gamma$  once  they are  detected by incoming experiments. 
\section*{Acknowledgments}
We thank Dr. Sumit Ghosh for useful discussions. We thank the referee for suggesting us many interesting constraints. The future experimental sensitivities may result in important correlations of the model parameters, which will be investigated in more detail in our future work. This research is funded by Vietnam Ministry of Education and Training and Hanoi Pedagogical University 2 under grant number B.2021-SP2-05. 

\appendix

\section{ \label{app_SMlikeHiggs} The SM-like Higgs boson}
For simplicity in estimating the SM-like Higgs boson $h$  found by LHC, we assume some conditions of the Higgs self-couplings  as follows:
\begin{align}\label{eq_SM-likeCond}
	\lambda_6 +\frac{f V}{2 t_{\beta} w}, \; \lambda_7 +\frac{f w}{2 t_{\beta}V}, \lambda_8 +\frac{f t_{\beta}V}{2 w}, \lambda_9 +\frac{f wt_{\beta}}{2 V} \simeq 0. 
\end{align}

As a result, the SM-like Higgs boson gets dominant contributions from the neutral Higgs basis $(\mathrm{Re}[\rho^0_2], \; \mathrm{Re}[\eta^0_1])$, corresponding to the following squared mass matrix:
\begin{align*}
	\mathcal{M}^2_h = 
	\begin{pmatrix}
		2 \lambda_2 c^2_{\beta} v^2 -\frac{ft_{\beta}Vw}{2}&   \lambda_5 s_{\beta}c_{\beta} v^2 +\frac{fVw}{2} \\ 
		\lambda_5 s_{\beta}c_{\beta} v^2 +\frac{fVw}{2}&   2\lambda_1 s^2_{\beta} v^2 -\frac{fVw}{2 t_{\beta}}
	\end{pmatrix},
\end{align*}
which  results in one light CP-even neutral Higgs boson with mass
\begin{equation}\label{eq_SMlikeh1}
	m_h^2= 2 \frac{ \mathcal{M}^2_{h,11}\mathcal{M}^2_{h,22} - \left(\mathcal{M}^2_{h,12}\right)^2}{\mathcal{M}^2_{h,11} +\mathcal{M}^2_{h,22} + \sqrt{\left( \mathcal{M}^2_{h,11} -\mathcal{M}^2_{h,22}\right)^2} +4\left(\mathcal{M}^2_{h,12}\right)^2}\varpropto \mathcal{O}(v^2).
\end{equation}
Similarly the 2HDM framework, this CP-even Higgs boson can be identified with the SM-like Higgs boson found from LHC. Denoting the mixing parameter $\alpha$ of these two Higgs bosons,
\begin{align}
	\begin{pmatrix}
		\mathrm{Re}[\rho^0_2]	\\
		\mathrm{Re}[\eta^0_1]	
	\end{pmatrix} & =  \begin{pmatrix}
		c_{\alpha}& s_{\alpha}	\\
		- s_{\alpha}	& c_{\alpha}
	\end{pmatrix}   \begin{pmatrix}
		H	\\
		h	
	\end{pmatrix},
\end{align}
the difference between the tree level couplings  $h\overline{e_a} e_a$ predicted by the model under consideration and the SM is a factor $s_{\alpha}/c_{\beta}$, exactly the same as  the 2HDM, see Ref. \cite{Nguyen:2020ehj} for example. Therefore, the combination of this factor and the loop corrections  can accommodate the experimental data.

\section{ \label{app_Zmm} One-loop corrections to $Z\mu^+\mu^-$ and $h\mu^+\mu^-$ vertices}
The relations between the flavor and physical base of the neutral gauge bosons is \cite{Long:2016lmj} 
\begin{equation}\label{eq_NG0}
	\begin{pmatrix}
		W_{3\mu}\\ 
		W_{8\mu}\\ 
		W_{15\mu}\\ 
		B''_{4\mu}
	\end{pmatrix} =
	\left(
	\begin{array}{cccc}
		s_W & c_W & 0 & 0 \\
		c_{32} c_W & -c_{32} s_W & -\, c_{\theta } s_{32} & s_{32} \, s_{\theta } \\
		c_{43} c_W s_{32} & -c_{43} s_{32} s_W & c_{32} c_{43} \, c_{\theta}-s_{43} \, s_{\theta } & -\, c_{\theta} s_{43}-c_{32} c_{43} \, s_{\theta} \\
		c_W s_{32} s_{43} & -s_{32} s_{43} s_W & c_{32} \, c_{\theta } s_{43}+c_{43} \, s_{\theta} & c_{43} \, c_{\theta }-c_{32} s_{43} \, s_{\theta } \\
	\end{array}
	\right)  \begin{pmatrix}
		A_{\mu}\\ 
		Z_{\mu}\\ 
		Z_{3\mu}\\ 
		Z_{4\mu}
	\end{pmatrix}, 
\end{equation}
where 
\begin{align}
	\label{eq_sijNG}
	s_{43}&= \frac{\sqrt{3- 6s^2_W}}{\sqrt{3 -4 s_W^2}} , \; c_{43}=-\sqrt{1-c_{43}^2},
	s_{32} = \frac{\sqrt{3- 4s^2_W}}{c_W}, \;  c_{32}= \sqrt{1-c_{32}^2},
	\crn   t_{2\theta}  &= \frac{s_{2\theta}}{c_{2\theta}} \sim \mathcal{O}(w^2/V^2), 
\end{align}
The matching relations from the breaking step $SU(4)_L \times U(1)_X\to SU(2)_L\times U(1)_Y$ is $W_3= W_3$, $B_{\mu}= c_{32}W_{8\mu}+ c_{43} s_{32}W_{15\mu} +s_{43} s_{32}B''_{\mu}$, and
\begin{equation}\label{eq_SMY}
	\frac{1}{\sqrt{3}}T_8  - \frac{2}{\sqrt{6}}T_{15} +X\mathtt{I}= \frac{Y}{2},
\end{equation}
where $B_{\mu}$ is the gauge boson of the gauge $U(1)_Y$ in the SM. The relation \eqref{eq_SMY} is consistent with the definition of the charge operator \eqref{eq_Qoperator}, the same as that given in the SM.   This means that we can consider $\eta$ and $\sigma$, $\nu_{aR}$ and $X_{aR}$ as new scalars and neutral fermions giving one-loop corrections to the $Z\mu^+\mu^-$ couplings as discussed in Ref. \cite{Kanemitsu:2012dc}. Because all new neutral fermions are singlets,  vertices $Z\nu_{aR} \bar{\nu}_{aR}$ and $ZX_{aR} \bar{X}_{aR}$ do not appear. Consequently, only diagrams 2 of Fig. 1 in Ref. \cite{Kanemitsu:2012dc} is irrelevant to our model.

We use the result from Ref. \cite{Kanemitsu:2012dc} having the following Yukawa couplings that give one-loop correction to the $g^{\mu}_{L,R}$ of the vertex $Z\mu^+\mu^-$:  
\begin{align*}
	\mathcal{L}_Y &= -y_L \bar{\mu}_L \phi_2  \chi_R -y_R  \bar{\mu}_R \phi \chi_L +\mathrm{h.c.},
\end{align*}
where $s_{k}$ and $\chi_{L,R}$ are physical Higgs and fermion states, and  $V_{k,l}$ is the Higg mixing parameters relating to $\phi$ and $\phi_2$ as follows: $\phi=V_{11}s_1 +V_{12}s_2$ and $\phi_2=V_{21}s_1 +V_{22}s_2$. 
The vertex corrections are:
\begin{align}
	\Delta g^{\mu}_L =& \frac{y^2_L}{16\pi^2} \left[ 2\sum_{k,l=1}^2\left\{ \left( -\frac{1}{2} -Q_s s^2_W\right) V^*_{2k} V_{2l} - Q_{s} s^2_WV^*_{1k} V_{1l} \right\} V_{2k} V^*_{2l} C_{24}(s_k,\chi,s_l; p,(q-p)) 
	\right.\crn &\left. - \sum_{k=1}^2 \left( -\frac{1}{2} +s^2_W\right) |V_{2k}|^2 (B_0 +B_1)(s_k,\chi;p)\right], 
	\crn	\Delta g^{\mu}_R =& \frac{y^2_R}{16\pi^2} \left[ 2\sum_{k,l=1}^2\left\{ \left( -\frac{1}{2} -Q_s s^2_W\right) V^*_{2k} V_{2l} - Q_{s} s^2_WV^*_{1k} V_{1l} \right\} V_{1k} V^*_{1l} C_{24}(s_k,\chi,s_l; p,(q-p)) 
	\right.\crn &\left. - \sum_{k=1}^2 s^2_W |V_{1k}|^2 (B_0 +B_1)(s_k,\chi;p)\right], 
\end{align}
where we have used $Q_{\chi}=0$ for neutral fermions, $C_{24}$ and $B_{0,1}$ are Passarino-Veltman (PV) functions, which transform into the notations of LoopTools (LT) \cite{Hahn:1998yk} as follows:
\begin{align}
	&\left\{ B_0, p^{\mu} B_1\right\}(A,B,p)\equiv 16\pi^2 \mu^{2\epsilon} \int \frac{d^nk}{i(2\pi)^n} \frac{\left\{1,\; k^{\mu}\right\}}{\left[k^2 -m^2_A +i\epsilon\right] \left[(k +p)^2 -m^2_B +i\epsilon\right]}
	\crn& \qquad \qquad\qquad\qquad\quad= \left\{ B_0, p^{\mu} B_1\right\}(p^2;m_A^2,m^2_B),
	\crn	& \left\{ p_1^{\mu} p_1^{\nu} C_{21} + p_2^{\mu} p_2^{\nu} C_{22} +(p_1^{\mu} p_2^{\nu}+ p_2^{\mu} p_1^{\nu}) C_{23} +g^{\mu \nu}C_{24}\right\}(A,B,C; p_1,p_2)
	\crn&=  16\pi^2 \mu^{2\epsilon} \int \frac{d^nk}{i(2\pi)^n} \frac{\left\{1,\; k^{\mu}\right\}}{\left[k^2 -m^2_A +i\epsilon\right] \left[(k +p_1)^2 -m^2_B +i\epsilon\right] +\left[(k +p_1+p_2)^2 -m^2_C +i\epsilon\right]}
	\crn &= \left\{ p_1^{\mu} p_1^{\nu} C_{11} + q^{\mu} q^{\nu} C_{22} +(p_1^{\mu} q^{\nu}+ q^{\mu} p_1^{\nu}) C_{12} +g^{\mu \nu}C_{00}\right\}(p_1^2,p_2^2, q^2;m^2_A,m^2_B,m^2_C) ,
\end{align}
where $B_{0,1}(p^2;m_A^2,m_B^2)$, and $C_{00,ij}(p_1^2,p_2^2, q^2;m^2_A,m^2_B,m^2_C) $ are LT notations, and $q=p_1+p_2$. In the particular case of the decay $Z\to \mu^+\mu^-$, we apply the on-shell conditions that $q^2=m_Z^2$ and $p_1^2=p_2^2=m^2_{\mu}\simeq 0$. 

The Yukawa part of  Eq. \eqref{eq_SSterm} gives the following equivalence $\eta^-_2 = \phi_2$ and  $\sigma^-\equiv \phi$ with $Q_{\phi}=Q_{\phi_2}=-1$, and new fermions are neutral singlets $\nu_{aR}, X_{aR}\sim (1,1,0)$, which do not couple to the $Z$ boson. The mixing parameter of Higgs bosons in Eq. \eqref{eq_scHigg} gives
\begin{equation} \label{eq_Higgsmix}
	-s_{\alpha} =V_{11}, \; c_{\alpha}= V_{12},\;  V_{21} = c_{\beta}c_{\alpha}, \; V_{22}=c_{\beta}s_{\alpha},  
\end{equation}
and $h^-_k=s_k$. 
The Yukawa couplings in Eq. \eqref{eq_L0ISSnumass} in the basis of new physical neutral fermions reads:
\begin{equation}\label{eq_Lychi}
	-\mathcal{ L}_Y= \sum_{i=1}^9 \sum_{a=1}^3 \left[ \frac{gM_0 \left(\hat{x}_{\nu}^{1/2} U^{\nu*}_{\mathrm{PMNS}}\right)_{a2}}{\sqrt{2} m_{W} s_{\beta}} U^{\nu*}_{(a+3)i}\bar{L}_{\mu} \eta n_{iR} + Y^{\sigma*}_{a2}U^{\nu}_{(a+6) i} \overline{\mu_R}n_{iL} \sigma^-\right] +\mathrm{h.c.}. 
\end{equation}
For a new fermion $n_{iL,R}$ with $Q_{n_i}=0$, we have
\begin{align*}
	y_{iL} &\equiv \sum_{a=1}^3 \frac{gM_0 \left(\hat{x}_{\nu}^{1/2} U^{\nu*}_{\mathrm{PMNS}}\right)_{a2}}{\sqrt{2} m_{W} s_{\beta}} U^{\nu*}_{(a+3)i}, \; y_{iR} \equiv  \sum_{a=1}^3 Y^{\sigma*}_{a2}U^{\nu}_{(a+6) i}, i=\overline{1,9}. 
\end{align*}
The vertex corrections to $Z{\mu^+\mu^-}$ in our work are:
\begin{align}
	\Delta g^{\mu}_L =&  \sum_{i=1}^9 \frac{|y_{iL}|^2}{16\pi^2}\left[ 2\sum_{k,l=1}^2\left\{ \left( -\frac{1}{2} +s^2_W\right) V^*_{2k} V_{2l} + s^2_WV^*_{1k} V_{1l} \right\} V_{2k} V^*_{2l} C_{00}
	\right.\crn &\left. - \sum_{k=1}^2 \left(- \frac{1}{2} +s^2_W\right) |V_{2k}|^2 (B_0 +B_1)\right]
	\crn&= \frac{g^2M_0^2}{32m_W^2 s^2_{\beta}} \left[U^{\dagger}_{\mathrm{PMNS}}\hat{x}_{\nu} U_{\mathrm{PMNS}}\right]_{22} \left[ 2\sum_{k,l=1}^2\left\{ \left( -\frac{1}{2} +s^2_W\right) V^*_{2k} V_{2l} + s^2_WV^*_{1k} V_{1l} \right\} V_{2k} V^*_{2l} C_{00}
	\right.\crn &\left. - \sum_{k=1}^2 \left( -\frac{1}{2} +s^2_W\right) |V_{2k}|^2 (B_0 +B_1)\right]
	%
	\crn	\Delta g^{\mu}_R =&  \sum_{i=1}^9 \frac{|y_{iR}|^2}{16\pi^2} \left[ 2\sum_{k,l=1}^2\left\{ \left( -\frac{1}{2} + s^2_W\right) V^*_{2k} V_{2l} + s^2_WV^*_{1k} V_{1l} \right\} V_{1k} V^*_{1l} C_{00}  
	\right.\crn &\left. - \sum_{k=1}^2 s^2_W |V_{1k}|^2 (B_0 +B_1) \right]
	\crn&=  \frac{|Y^{\sigma \dagger}Y^{\sigma}|_{22}}{16\pi^2} \left[ 2\sum_{k,l=1}^2\left\{ \left( -\frac{1}{2} + s^2_W\right) V^*_{2k} V_{2l} + s^2_WV^*_{1k} V_{1l} \right\} V_{1k} V^*_{1l} C_{00}  
	\right.\crn &\left. - \sum_{k=1}^2 s^2_W |V_{1k}|^2 (B_0 +B_1) \right],
\end{align}
where $C_{00}= C_{00}(0,0,m_Z^2; m^2_{h^\pm_k}, M_0^2,m^2_{h^\pm_l})$ and $B_{0,1}=B_{0,1}(0,m^2_{h_k},M_0^2)$. 

The following modified $Z \mu^+ \mu^-$ couplings are 
\begin{equation}
	i\frac{g}{c_W}\gamma_{\mu}\left[ \left(g^{\mathrm{SM},\mu}_L + \Delta g^{\mu}_L\right)P_L + \left(g^{\mathrm{SM},\mu}_R + \Delta g^{\mu}_R\right) P_R\right],
\end{equation}
where $g^{\mathrm{SM},\mu}_L=-1/2 +s_W^2$, and  $g^{\mathrm{SM},\mu}_R=s_W^2$.
Defining the following quantity \cite{Escribano:2021css}:
\begin{equation}
	\label{eq_dRZll}
	\delta R_{Ze_ae_a}\equiv \frac{\Gamma(Z\to e_a^+e_a^-)}{\Gamma_{\mathrm{SM}}(Z\to e_a^+e_a^-)}-1= \frac{\left|g_V\right| ^2 +\left|g_A\right| ^2}{\left|g^{\mathrm{SM}}_V\right| ^2 +\left|g^{\mathrm{SM}}_Z\right| ^2} -1,
\end{equation}
where $g^{\mathrm{SM}}_{V,A}=g^{\mathrm{SM}}_R\pm g^{\mathrm{SM}}_L$, $g_V=g^{\mathrm{SM}}_{V}+\Delta g^{\mu}_L+ \Delta g^{\mu}_R$, and  $g_A= g^{\mathrm{SM}}_{A}- \Delta g^{\mu}_L+ \Delta g^{\mu}_R$, 
the experimental constraint  is :  $ -7< \delta  R_{Ze_ae_a} \times 10^{3}<6$ \cite{Escribano:2021css, ParticleDataGroup:2020ssz}.

\end{document}